\documentclass[aps,preprint,pra,showpacs,amsfonts,amssymb,amsmath,nofootinbib]{revtex4-1}

\usepackage{subfig}
\usepackage{dsfont}
\usepackage{appendix}
\usepackage{color}
\usepackage{graphicx}
\usepackage{bbold}

\begin{document}

%%%%%%%%%%%%%%%%%%%%%%%%%%%%%%%%%%%%%%%%%%%%%%%%%%%%%%%%%%%
\title{\bf Integral method for the calculation of Hawking radiation in dispersive media\\
II. Asymmetric asymptotics}

\author{Scott Robertson}
\affiliation{Laboratoire de Physique Th\'{e}orique, CNRS UMR 8627,\\
B\^{a}timent 210, Universit\'{e} Paris-Sud 11, 91405 Orsay Cedex, France}
\email{scott.robertson@th.u-psud.fr}

\begin{abstract}
Analogue gravity experiments make feasible the realisation of black hole spacetimes in a laboratory setting and the observational verification of Hawking radiation.  Since such analogue systems are typically dominated by dispersion, efficient techniques for calculating the predicted Hawking spectrum in the presence of strong dispersion are required.  In the preceding paper, an integral method in Fourier space is proposed for stationary $1+1$-dimensional backgrounds which are asymptotically symmetric.  Here, this method is generalised to backgrounds which are different in the asymptotic regions to the left and right of the scattering region.
\end{abstract}


\maketitle

\newpage

%%%%%%%%%%%%%%%%%%%%%%%%%%%%%%%%%%%%%%%%%%%%%%%%%%%%%%%%

\section{Introduction
\label{sec:Introduction}}

Observational verification of Hawking radiation \cite{Hawking-1974,Hawking-1975} is within reach thanks to experimentally accessible analogues of gravity that mimic the behaviour of wave propagation in a black hole spacetime \cite{ArtificialBlackHoles,Schutzhold-Unruh,LivingReview}.  Unruh's original proposal of acoustic waves in a flowing fluid that crosses the speed of sound \cite{Unruh-1981} has since been expanded upon and generalised, resulting in a cornucopia of analogue gravity systems that includes water waves \cite{Schutzhold-Unruh-2002,Rousseaux-et-al-2008,Rousseaux-et-al-2010,Weinfurtner-et-al-2011}, light in nonlinear media \cite{Philbin-et-al-2008,Belgiorno-et-al-2010}, and phononic excitations in atomic BEC \cite{Garay-et-al-2000,Garay-et-al-2001,Barcelo-Liberati-Visser-2001-arXiv,Barcelo-Liberati-Visser-2001} and in quantum fluids of light \cite{Solnyshkov-et-al-2011,Gerace-Carusotto-2012}.  As for the gravitational black hole, these analogue systems induce scattering between waves of {\it opposite norm} (see, {\it e.g.}, \cite{Robertson-2012}, and also Part I).  Since the sign of the norm of a field mode indicates whether its amplitude becomes an annihilation or a creation operator upon quantization of the field \cite{Robertson-2012}, scattering between waves of opposite norm leads to a mixing of annihilation and creation operators in the basis transformation between ingoing and outgoing waves, and hence to the inequality of the ingoing and outgoing vacuum states.  This is the essence of Hawking radiation: in the absence of ingoing particles, outgoing particles must be present, so {\it particles are emitted spontaneously.}  The spectrum of the spontaneous radiation is directly related to the scattering amplitudes between waves of opposite norm (see Part I), so the Hawking spectrum is determined by the scattering properties of the analogue spacetime.

Unlike classical gravity, analogue gravity systems are typically dominated by dispersion \cite{Jacobson-1991}, complicating the wave behaviour and making it less amenable to analytical techniques.  Existing techniques include FDTD wave propagation \cite{Unruh-1995}, which is computationally intensive, especially if a spectrum over a wide range of frequencies is required; numerical solution of an ODE at fixed frequency \cite{Corley-Jacobson-1996,Macher-Parentani-2009}, which can be performed if the dispersion relation is a polynomial of low degree and any exponentially growing waves do not significantly affect the accuracy of the solution; analytical solution of the ODE for a step discontinuous background by matching the solutions at the discontinuity \cite{Corley-1997,Recati-et-al-2009,Finazzi-Parentani-2012}; and analytical techniques which are valid when the background varies slowly over length scales at which dispersive effects become important \cite{Corley-1998,Leonhardt-Robertson-2012}.  However, practical setups will often lie outside these domains of applicability: dispersion relations can be too complicated to be modelled by a low-degree polynomial, and the background may have to vary rapidly in order to boost the spectrum to an observable level.  Our aim is to have a method for calculating the scattering amplitudes which relaxes the restrictions currently in place.

In the preceding paper -- referred to here as Part I -- a numerical method is described which treats the wave equation as an integral equation in Fourier space.  This has the advantage that the dispersion relation appears as a multiplicative function (rather than a differential operator), and can be quite arbitrary.  Through discretisation of the integral, the wave equation is transformed into a linear equation with the integral kernel as a matrix, and this can be solved using standard and efficient numerical algorithms.  The one restriction imposed in Part I is that the background be asymptotically symmetric -- that is, the effective ``spacetime'', which is assumed to approach a limiting value so that ingoing and outgoing waves are well-defined, is the same in the left- and right-hand asymptotic regions.  While this is not an uncommon situation (it applies to nonlinear optical analogues \cite{Philbin-et-al-2008,Belgiorno-et-al-2010}, for example), it is far from the most general case.  The purpose of the present paper is to remove this restriction: to generalise the integral method to backgrounds that asymptote to {\it different} values in the left- and right-hand asymptotic regions.

This paper is organised as follows.  In Section \ref{sec:Asymmetric_integral_equation}, we briefly review some aspects already covered in Part I, particularly how the wave equation manifests itself as an integral equation in Fourier space and how decomposition into left- and right-Fourier transforms allows us to exploit certain analyticity properties of the components of the solution.  We shall also note the essential differences induced by asymmetry, and how these complicate the method.  In Section \ref{sec:Step-discontinuous_background}, we present the solution of the integral equation for the simplest asymmetric background: that which is homogeneous except for a step discontinuity.  This introduces the new mathematical machinery required to deal with asymmetry, stripped of the additional details due to the precise nature of the variation of the background; these additional details are then replaced in Section \ref{sec:General_background}.  Section \ref{sec:Application} looks at the application of the method to a simple concrete model, and the paper concludes with Section \ref{sec:Conclusion}.

%%%%%%%%%%%%%%%%%%%%%%%%%%%%%%%%%%%%%%%%%%%%%%%%%%%%%%%%

\section{Asymmetric integral equation
\label{sec:Asymmetric_integral_equation}}

Here we shall briefly review the form of the wave equation as an integral equation in Fourier space, the analytic manipulations required to make it soluble, and the differences that arise due to asymmetry of the asymptotic regions.

%%%%%%%%%%%%%%%%%%%%%%%%%%%%

\subsection{Wave equation as integral equation
\label{sub:Wave_equation_as_integral_equation}}

While it is to be emphasised that the integral method is applicable to a wide variety of wave equations, we shall for definiteness focus our attention on Unruh's acoustic model with dispersion \cite{Unruh-1995}.  With $c(k)$ the wavevector-dependent speed of sound, $u(x)$ the position-dependent flow velocity and $\phi(x,t)=\phi_{\omega}(x)\,e^{-i \omega t}$ a stationary wave solution, the wave equation is
\begin{equation}
\left[ \left(-i\omega + \partial_{x}u(x) \right) \left(-i\omega + u(x)\partial_{x}\right) - c^{2}\left(-i\partial_{x}\right) \partial_{x}^{2} \right] \phi_{\omega}(x) = 0 \,.
\label{eq:wave_eqn}
\end{equation}
Instead of the solution in position space, $\phi_{\omega}(x)$, we shall consider the Fourier-transformed solution
\begin{equation}
\psi_{\omega}(k) = \int_{-\infty}^{+\infty} \mathrm{d}x \, e^{-ikx} \, \phi_{\omega}(x) \,.
\label{eq:wave_FT}
\end{equation}
Fourier transforming Eq. (\ref{eq:wave_eqn}), we find an equivalent equation for $\psi_{\omega}(k)$:
\begin{equation}
g_{\omega}(k) \, \psi_{\omega}(k) + \int_{-\infty}^{+\infty} \mathrm{d}k^{\prime} \, K_{\omega}(k,k^{\prime}) \, \psi_{\omega}(k^{\prime}) = 0 \,,
\label{eq:wave_eqn_integral}
\end{equation}
where $g_{\omega}(k)$ encodes the position-independent part of Eq. (\ref{eq:wave_eqn}) that contains only constants and derivatives, while $K(k,k^{\prime})$ encodes the position-dependent terms which appear as convolutions of Fourier transforms.  Explicitly, using $\mathcal{F}[f](k)$ to denote the Fourier transform of an arbitrary function $f(x)$, we have
\begin{eqnarray}
g_{\omega}(k) & = & c^{2}(k)k^{2} - \omega^{2} \,,
\label{eq:unruh_dispersion} \\
K_{\omega}(k,k^{\prime}) & = & \frac{1}{2\pi} \left[2 \, \omega \, k \, \mathcal{F}[u](k-k^{\prime}) + i\omega \, \mathcal{F}[\partial_{x}u](k-k^{\prime}) \right. \nonumber \\
& & \qquad \qquad \left. - k^{2} \, \mathcal{F}[u^{2}](k-k^{\prime}) - ik \, \mathcal{F}[\partial_{x}u^{2}](k-k^{\prime}) \right] \nonumber \\
& = & \frac{1}{2\pi} \left[ \omega \, (k+k^{\prime}) \, \mathcal{F}[u](k-k^{\prime}) - k \, k^{\prime} \, \mathcal{F}[u^{2}](k-k^{\prime}) \right] \,,
\label{eq:unruh_kernel}
\end{eqnarray}
where the second line of Eq. (\ref{eq:unruh_kernel}) follows from the relation $\mathcal{F}[\partial_{x}u](k) = ik\,\mathcal{F}[u](k)$.  To avoid cumbersome expressions, we shall suppress from now on the explicit dependence of the various quantities on the frequency $\omega$.

%%%%%%%%%%%%%%%%%%%%%%%%%%%%

\subsection{Left- and right-Fourier transforms
\label{sub:Left_and_right_Fourier_transforms}}

By considering {\it half}-Fourier transforms of the field, we can decompose it into two parts, as follows:
\begin{alignat}{2}
\psi^{L}(k) = \int_{-\infty}^{0} \mathrm{d}x \, e^{-ikx} \, \phi(x) \,, & \qquad \psi^{R}(k) = \int_{0}^{+\infty} \mathrm{d}x \, e^{-ikx} \, \phi(x) \,,
\label{eq:half-FTs}
\end{alignat}
so that
\begin{equation}
\psi(k) = \psi^{L}(k) + \psi^{R}(k)\,.
\label{eq:field_decomposition}
\end{equation}
Since the field $\phi(x)$ is assumed to be asymptotically bounded, $\psi^{L}(k)$ is analytic and goes to zero at least as fast as $1/k$ in the upper half complex $k$-plane, while the same is true of $\psi^{R}(k)$ in the lower half complex $k$-plane.

The integral kernel $K(k,k^{\prime})$ can be similarly decomposed into functions with special analyticity properties in the integration variable $k^{\prime}$:
\begin{equation}
K(k,k^{\prime}) = K_{L}(k,k^{\prime}) + K_{R}(k,k^{\prime}) + K_{\mathrm{step}}(k) \,.
\label{eq:kernel_decomposition}
\end{equation}
Here, $K_{L}(k,k^{\prime})$ and $K_{R}(k,k^{\prime})$ are analytic and vanish asymptotically in the {\it lower} and {\it upper} half complex $k^{\prime}$-planes, respectively, and as in Part I this difference is indicated by the appearance of $L$ and $R$ as subscripts rather than superscripts.  With the $k^{\prime}$-dependence contained entirely in the arguments of Fourier transforms, as in the first line of Eq. (\ref{eq:unruh_kernel}), $K_{L}(k,k^{\prime})$ and $K_{R}(k,k^{\prime})$ are formed simply by replacing the full Fourier transform with left and right-Fourier transforms, respectively.  $K_{\mathrm{step}}(k)$ is independent of $k^{\prime}$; its inclusion allows us to deal with step discontinuities at $x=0$ (as mentioned in Appendix A of Part I).

If two functions are analytic on the same half-plane, then the integral of their product is the same for any deformation of the integration contour onto this half-plane; furthermore, if each of these functions behaves asymptotically like $1/k$, then the integral of their product vanishes.  Substituting $\psi(k)$ and $K(k,k^{\prime})$ in Eq. (\ref{eq:wave_eqn_integral}) for their decompositions in Eqs. (\ref{eq:field_decomposition}) and (\ref{eq:kernel_decomposition}), this means that the integrals of $K_{R}(k,k^{\prime}) \cdot \psi^{L}(k^{\prime})$ and $K_{L}(k,k^{\prime}) \cdot \psi^{R}(k^{\prime})$ must vanish, and the integral equation becomes
\begin{multline}
g(k) \left(\psi^{L}(k) + \psi^{R}(k) \right) + \int_{-\infty}^{+\infty} \mathrm{d}k^{\prime} \, K_{L}(k,k^{\prime}) \psi^{L}(k^{\prime}) + \int_{-\infty}^{+\infty} \mathrm{d}k^{\prime} \, K_{R}(k,k^{\prime}) \psi^{R}(k^{\prime}) \\
+ K_{\mathrm{step}}(k) \, \int_{-\infty}^{+\infty} \mathrm{d}k^{\prime} \left( \psi^{L}(k^{\prime}) + \psi^{R}(k^{\prime}) \right) = 0 \,.
\label{eq:int_eqn_field_kernel_decomp}
\end{multline}

%%%%%%%%%%%%%%%%%%%%%%%%%%%%

\subsection{Extracting the asymptotic dispersion relations
\label{sub:Extracting_the_asymptotic_dispersion_relations}}

In order that ingoing and outgoing waves are well-defined, we assume that the background $u(x)$ asymptotes to a constant value in both the left- and right-hand regions:
\begin{equation}
u(x) \rightarrow \begin{cases} u_{L} & \mbox{as } x \rightarrow -\infty \\
u_{R} & \mbox{as } x \rightarrow +\infty \end{cases} .
\end{equation}
These constant values contribute singular terms to the half-Fourier transforms of $u$: introducing the label $\sigma \in \left\{L,R\right\}$ and the sign $s_{L}=-1$ and $s_{R}=1$, we have
\begin{equation}
\mathcal{F}_{\sigma}[u](k) = u_{\sigma} \left[ \pi\,\delta(k) + \frac{s_{\sigma}}{i\,k} \right] + \mathcal{F}_{\sigma}[u-u_{\sigma}](k) \,,
\label{eq:FTu_singular}
\end{equation}
where the asymptotic vanishing of $u-u_{\sigma}$ in the integrated region causes the second term to be regular in $k$.  A similar expression holds for $\mathcal{F}_{\sigma}[u^{2}]$, while derivatives of the background, such as appear in Eq. (\ref{eq:unruh_kernel}), can have no singular component since they must vanish asymptotically.  These singular terms are thus contained in the kernels\footnote{This does not contradict the half-plane analyticity of $K_{\sigma}(k,k^{\prime})$, since the singularities occur on the real line, which is excluded from both half-planes.  Analyticity on a half-plane comes from asymptotic {\it boundedness} of the waveform $\phi(x)$, while singularities on the real line can occur because $\phi(x)$ need not be asymptotically {\it vanishing}.} $K_{\sigma}(k,k^{\prime})$, and can be immediately integrated in Eq. (\ref{eq:int_eqn_field_kernel_decomp}), returning a factor times $\psi^{\sigma}(k)$ and leaving residual kernels which are smooth functions of $k$ and $k^{\prime}$.  When $u_{L}=u_{R}$ -- the case studied in our previous paper -- this leaves the integral in the same form as Eq. (\ref{eq:wave_eqn_integral}) or Eq. (\ref{eq:int_eqn_field_kernel_decomp}), with $g(k)$ and $K(k,k^{\prime})$ appropriately redefined.  However, when $u_{L} \neq u_{R}$, the factors multiplying $\psi^{L}(k)$ and $\psi^{R}(k)$ are different.  This results in two different coefficient functions $g_{L}(k)$ and $g_{R}(k)$, so that Eq. (\ref{eq:int_eqn_field_kernel_decomp}) now takes the form
\begin{multline}
g_{L}(k) \psi^{L}(k) + g_{R}(k) \psi^{R}(k) + \int_{-\infty}^{+\infty} \mathrm{d}k^{\prime} \, K_{L}(k,k^{\prime}) \psi^{L}(k^{\prime}) + \int_{-\infty}^{+\infty} \mathrm{d}k^{\prime} \, K_{R}(k,k^{\prime}) \psi^{R}(k^{\prime}) \\
+ K_{\mathrm{step}}(k) \, \int_{-\infty}^{+\infty} \mathrm{d}k^{\prime} \, \left( \psi^{L}(k^{\prime}) + \psi^{R}(k^{\prime}) \right) = 0 \,,
\label{eq:int_eqn_diff_asymptotics}
\end{multline}
where
\begin{equation}
g_{\sigma}(k) = c^{2}(k)k^{2} - \left( \omega - u_{\sigma}k \right)^{2} \,.
\label{eq:left-right_dispersion}
\end{equation}
The roots of Eq. (\ref{eq:left-right_dispersion}) are the solutions of the dispersion relation in the asymptotic regions: it is only where $g_{\sigma}(k)=0$ that $\psi^{\sigma}(k)$ can be singular.

%%%%%%%%%%%%%%%%%%%%%%%%%%%%

\subsection{Complications due to asymmetry
\label{sub:Complications_due_to_asymmetry}}

Equation (\ref{eq:int_eqn_diff_asymptotics}) is the analogue, for a profile with asymmetric asymptotic regions, of Eq. (18) of Part I.  Apart from the inclusion of $K_{\mathrm{step}}(k)$ -- which could have been included in Part I, but was set to zero as we considered only backgrounds continuous at $x=0$ -- the only difference between these equations is the inequality of $g_{L}(k)$ and $g_{R}(k)$ in the asymmetric case.  This apparently straightforward generalization requires a considerable amount of additional machinery to extract the scattering matrix in a similar manner to Part I.  There, we noted that, while the vanishing of $g(k)$ at certain points on the real axis made the integral operator singular, it could be transformed into an invertible operator by an appropriate regularisation procedure \cite{Bart-Warnock-1973}, in which the regular part of the solution was isolated and found to obey a regularity condition for each zero of $g(k)$; these could then be subtracted from the integral equation in such a way as to allow division by $g(k)$, leaving behind a non-singular equation.  However, in Eq. (\ref{eq:int_eqn_diff_asymptotics}), the coefficient functions $g_{L}(k)$ and $g_{R}(k)$ have different roots; there is thus a ``doubling'' of the number of singularities\footnote{Not necessarily a strict doubling, as $g_{L}(k)$ and $g_{R}(k)$ may have different numbers of real roots.} in $\psi(k)$, split between $\psi^{L}(k)$ and $\psi^{R}(k)$, which are now singular at different points.  While we can still isolate the regular parts of the solution and define a regularity condition at each of the zeros of $g_{L}(k)$ and $g_{R}(k)$, the fact that these zeros are different means that regularising Eq. (12) such that one of the two coefficient functions can be divided out still leaves the other intact, leaving coupled integral equations that cannot be solved directly.

To deal with this, we first split the integral equation into two parts, each containing only one of the coefficient functions $g_{\sigma}(k)$ and each of which can be regularised according to the procedure of Part I.  This splitting of the integral equation constitutes the additional machinery required in the asymmetric case, and it is effected by exploiting the analyticity properties of the $\psi^{\sigma}(k)$ described above.  It is inspired by a similar method used to solve the {\it Riemann-Hilbert problem} \cite{Estrada-Kanwal-1987}, to which the integral equation reduces in the case of a piecewise homogeneous background with a step discontinuity.  In Section \ref{sec:Step-discontinuous_background}, attention is focused on this case in order to acquaint ourselves with the new machinery; the generalisation to arbitrary velocity profiles is described in Section \ref{sec:General_background}, unifying the Riemann-Hilbert elements introduced in Section \ref{sec:Step-discontinuous_background} with the non-trivial integral kernels considered in Part I.

There is one final caveat to be mentioned here: we shall restrict ourselves to coefficient functions $g_{\sigma}(k)$ -- and equivalently to dispersion relations $c^{2}(k)$ -- which are {\it polynomials}, and whose roots we know or can calculate efficiently.  This is because, in the new machinery inherited from the Riemann-Hilbert problem, we shall require some analytic manipulations of the $g_{\sigma}(k)$, similar to the splitting of the solution and the kernel into half-plane analytic parts.  For general $g_{\sigma}(k)$, this is non-trivial and computationally intensive; for polynomials, by contrast, many of these manipulations can be done ``by hand''.  We shall not need to make any restrictions on the degree of these polynomials -- in particular, we are {\it not} reintroducing the restriction to low-degree polynomial dispersion relations required for stable numerical solution of ODEs.  Indeed, given an arbitrary dispersion relation, we can find a polynomial approximation to it over any finite range of $k$; so long as the wavevectors of interest are contained within this range, and no additional real roots are generated outside this range, we expect this polynomial approximation to return a very good approximation to the scattering amplitudes.

%%%%%%%%%%%%%%%%%%%%%%%%%%%%%%%%%%%%%%%%%%%%%%%%%%%

\section{Step-discontinuous background
\label{sec:Step-discontinuous_background}}

The simplest situation with asymmetric asymptotic regions in that in which the velocity profile is uniform everywhere except for a step discontinuity.  It will be instructive to first consider this case, for its solution deals with the essential complications arising from asymptotic asymmetry.  Generalisation to arbitrary profiles involves unifying this method with the inversion of non-trivial integral kernels described in Part I, and will be treated in Section \ref{sec:General_background}.

%%%%%%%%%%%%%%%%%%%%%%%%%%%%

\subsection{Riemann-Hilbert problem
\label{sub:Riemann-Hilbert_problem}}

Adopting a step-discontinuous profile amounts, at the level of the integral equation (\ref{eq:int_eqn_diff_asymptotics}), to setting $K_{L}$ and $K_{R}$ to zero, while $K_{\mathrm{step}}(k)$, which is that part of the kernel (\ref{eq:unruh_kernel}) coming from the (singular) derivatives of $u$ and $u^{2}$, is non-zero:
\begin{equation}
g_{L}(k) \, \psi^{L}(k) + g_{R}(k) \, \psi^{R}(k) + \phi_{0} \, K_{\mathrm{step}}(k) = 0 \,,
\label{eq:step_RH}
\end{equation}
where
\begin{equation}
K_{\mathrm{step}}(k) = \frac{1}{2\pi i} \left( u_{R}^{2} - u_{L}^{2} \right) \left( k - \frac{\omega}{u_{R}+u_{L}} \right) \,
\label{eq:K_step}
\end{equation}
and where we have defined
\begin{equation}
\phi_{0} = 2\pi \, \phi(x=0) = \int_{-\infty}^{+\infty} \mathrm{d}k^{\prime} \left( \psi^{L}(k^{\prime}) + \psi^{R}(k^{\prime}) \right) \,.
\label{eq:phi0_defn}
\end{equation}
Noting that $\phi_{0}$ is just a number, Eq. (\ref{eq:step_RH}) is a {\it non-normal inhomogeneous Riemann-Hilbert problem} \cite{Estrada-Kanwal-1987}: a functional equation between the boundary values of two functions analytic in neighbouring regions of the complex plane, which in this case are the upper and lower half-planes, the boundary being the real axis.  It is {\it inhomogeneous} because it contains a term not proportional to $\psi^{\sigma}(k)$.  More importantly, it is {\it non-normal} because the coefficient functions $g_{\sigma}(k)$ vanish at discrete points on the real axis, allowing the solutions $\psi^{\sigma}(k)$ to be singular at these points.  As for integral equations, this renders the equation non-invertible and the solutions non-unique, and due to linearity\footnote{Equation (\ref{eq:step_RH}) is linear in the sense that $\phi_{0}$, as defined in Eq. (\ref{eq:phi0_defn}), scales linearly with $\psi$.  If instead $\phi_{0}$ is considered as fixed, the equation is no longer linear and the solutions are restricted to a hypersurface rather than spanning a whole vector space.} of the equation, there is a vector space of solutions.  As mentioned in Part I, the dimension $N$ of this vector space is the number of ingoing or outgoing asymptotic wavevector solutions, equal to half of the total number of (real) asymptotic wavevector solutions.

%%%%%%%%%%%%%%%%%%%%%%%%%%%%

\subsection{Splitting into half-plane analytic parts
\label{sub:Splitting_into_half-plane_analytic_parts}}

It was expressed in \S\ref{sub:Complications_due_to_asymmetry} that the functions $g_{\sigma}(k)$ shall be restricted to polynomials whose roots are known.  For a dispersive medium, $c^{2}(k)$ is then a polynomial of even degree larger than $2$, and it thus follows that the coefficient of the highest power of $k$ -- call it $G_{0}$ -- is the same for both $g_{L}(k)$ and $g_{R}(k)$.  Moreover, we shall assume that the medium is non-dissipative in the asymptotic regions, so that incoming waves can be incident from infinity; this implies that the $g_{\sigma}(k)$ are {\it real} polynomials, and that complex roots must occur in complex conjugate pairs.  Therefore, we can write
\begin{equation}
g_{\sigma}(k) = G_{0} \, \prod_{j=1}^{N^{r}_{\sigma}} \left( k - k^{r}_{\sigma,j} \right) \, \prod_{q=1}^{N^{c}_{\sigma}} \left( k - k^{c}_{\sigma,q} \right) \left( k - k^{c\star}_{\sigma,q} \right) \,.
\label{eq:g_expansion}
\end{equation}
Here, there are $N^{r}_{\sigma}$ real roots $k_{\sigma,j}^{r}$ of $g_{\sigma}(k)$, and $N_{\sigma}^{c}$ pairs of complex roots $\left(k_{\sigma,q}^{c},k_{\sigma,q}^{c\star}\right)$.  The degree of $g_{\sigma}(k)$ is $N_{\sigma}^{r}+2N_{\sigma}^{c}$, and is independent of $\sigma$.

The roots of $g_{\sigma}(k)$ are the solutions of the dispersion in asymptotic region $\sigma$, so that the general solution there is a sum of plane waves of the form $\mathrm{exp}\left(i k_{i} x\right)$.  But physical solutions must be asymptotically bounded, so that certain of these plane waves are physically forbidden.  All real wavevectors are allowed (though they can be separated into ingoing and outgoing waves according to the signs of their group velocities -- see \S II of Part I), while only those complex wavevectors that generate asymptotically vanishing evanescent waves are allowed; these have $\mathrm{Im}\left[k^{c}_{L}\right] < 0$ on the left-hand side, and $\mathrm{Im}\left[k^{c}_{R}\right] > 0$ on the right-hand side.  For definiteness, we define $k_{\sigma,q}^{c}$ to be the allowed wavevectors in asymptotic region $\sigma$, while $k_{\sigma,q}^{c\star}$ will be the exponentially divergent forbidden wavevectors.

We shall soon see that the analyticity of the first and second terms of Eq. (\ref{eq:step_RH}) on the upper and lower half planes, respectively, allows the sought-for splitting of the equation, while the asymptotic divergence of the equation determines the number of degrees of freedom in the solution.  It is possible to tame the asymptotic divergence -- zooming in on the actual degrees of freedom -- while leaving the analyticity of the first two terms in tact.  This is done by dividing out all of the factors $k-k^{c\star}_{\sigma,j}$, where the $k^{c\star}_{\sigma,j}$ are the forbidden wavevectors; it leaves us with the equivalent equation
\begin{equation}
\tilde{g}_{L}(k) \, \psi^{L}(k) + \tilde{g}_{R}(k) \, \psi^{R}(k) + \phi_{0} \, \kappa(k) = 0 \,,
\label{eq:step_RH_kappa_unsplit}
\end{equation}
where we have defined
\begin{eqnarray}
\tilde{g}_{\sigma}(k) & = & G_{0} \, \prod_{j=1}^{N_{\sigma}^{r}} \left( k - k_{\sigma,j}^{r} \right) \frac{\prod_{q=1}^{N_{\sigma}^{c}} \left( k - k_{\sigma,q}^{c} \right)}{\prod_{q=1}^{N_{-\sigma}^{c}} \left( k - k_{-\sigma,q}^{c\star} \right)} \,,
\label{eq:g_tilde_defn} \\
\kappa(k) & = & \frac{K_{\mathrm{step}}(k)}{\prod_{q=1}^{N_{L}^{c}}\left(k-k_{L,q}^{c\star}\right) \, \prod_{q=1}^{N_{R}^{c}} \left(k-k_{R,q}^{c\star}\right)} \,,
\label{eq:kappa_defn}
\end{eqnarray}
and where by $-\sigma$ we mean the opposite asymptotic region from that denoted by $\sigma$.  Equation (\ref{eq:step_RH_kappa_unsplit}) is thus split into half-plane analytic parts if $\kappa(k)$ is itself split in this way.  Indeed, since $K_{\mathrm{step}}(k)$ is a first-degree polynomial, $\kappa(k)$ is exactly a sum of poles, one at each of the forbidden wavevectors.  Collecting those on the lower half plane together gives that part of $\kappa(k)$ which is analytic on the upper half plane, and vice versa.  Therefore, we can write
\begin{equation}
\kappa(k) = \kappa^{L}(k) + \kappa^{R}(k)
\label{eq:kappa_LRsplit}
\end{equation}
where
\begin{subequations}
\begin{alignat}{2}
\kappa^{L}(k) = \sum_{p=1}^{N_{R}^{c}} C_{R,p}^{-1} \, \frac{K_{\mathrm{step}}\left(k_{R,p}^{c\star}\right)}{k-k_{R,p}^{c\star}} \, , & \qquad C_{R,p} = \prod_{q=1}^{N_{L}^{c}} \left( k_{R,p}^{c\star} - k_{L,q}^{c\star} \right) \, \prod_{q=1,q \neq p}^{N_{R}^{c}} \left(k_{R,p}^{c\star} - k_{R,q}^{c\star} \right) \, ,\\
\kappa^{R}(k) = \sum_{p=1}^{N_{L}^{c}} C_{L,p}^{-1} \, \frac{K_{\mathrm{step}}\left(k_{L,p}^{c\star}\right)}{k-k_{L,p}^{c\star}} \, , & \qquad C_{L,p} = \prod_{q=1,q \neq p}^{N_{L}^{c}} \left( k_{L,p}^{c\star} - k_{L,q}^{c\star} \right) \, \prod_{q=1}^{N_{R}^{c}} \left(k_{L,p}^{c\star} - k_{R,q}^{c\star} \right) \,.
\end{alignat}
\label{eq:kappa_split}\end{subequations}
The superscript on $\kappa^{\sigma}(k)$ indicates that it is analytic on the same half plane as $\psi^{\sigma}(k)$.

Finally, we can rewrite Eq. (\ref{eq:step_RH_kappa_unsplit}) in the form
\begin{equation}
\tilde{g}_{L}(k) \, \psi^{L}(k) + \phi_{0} \, \kappa^{L}(k) = -\tilde{g}_{R}(k) \, \psi^{R}(k) - \phi_{0} \, \kappa^{R}(k) \,.
\label{eq:step_RH_analytic}
\end{equation}
The left-hand side of Eq. (\ref{eq:step_RH_analytic}) is manifestly analytic on the upper half plane, and the right-hand side on the lower half plane.  Both sides are thus equal to an {\it entire} function ({\it i.e.} one that is analytic everywhere).  This entire function is restricted by the asymptotic behaviour of Eq. (\ref{eq:step_RH_analytic}) as $\left| k \right| \rightarrow \infty$.  We have noted that $\psi^{\sigma}(k)$ behaves asymptotically like $k^{-1}$, and it is clear from (\ref{eq:kappa_split}) that $\kappa^{\sigma}(k)$ behaves similarly.  From Eq. (\ref{eq:g_tilde_defn}) we see that $\tilde{g}_{\sigma}(k) \sim k^{\tilde{d}}$ where $\tilde{d} = N_{L}^{r}+\left(N_{L}^{c}-N_{R}^{c}\right)/2 = N_{R}^{r}+\left(N_{R}^{c}-N_{L}^{c}\right)/2 = \left(N_{L}^{r}+N_{R}^{r}\right)/2 = N$, which is the number of ingoing or outgoing waves.  As a whole, then, Eq. (\ref{eq:step_RH_analytic}) behaves asymptotically like $k^{N-1}$, and the entire function to which it is equal behaves in the same way.  The only entire function with the correct asymptotic behaviour is a polynomial of degree $N-1$, which has $N$ degrees of freedom -- exactly the number of degrees of freedom in the space of (asymptotically bounded) solutions.  Then, in place of Eq. (\ref{eq:step_RH_analytic}), we can write two equations:
\begin{subequations}
\begin{eqnarray}
\tilde{g}_{L}(k) \, \psi^{L}(k) + \phi_{0} \, \kappa^{L}(k) & = & \sum_{j=1}^{N} P_{j} \, p_{j}(k) \,, \\
\tilde{g}_{R}(k) \, \psi^{R}(k) + \phi_{0} \, \kappa^{R}(k) & = & - \sum_{j=1}^{N} P_{j} \, p_{j}(k) \,,
\end{eqnarray}
\label{eq:step_RH_split}\end{subequations}
where the $p_{j}(k)$ are an arbitrarily chosen set of $N$ linearly independent polynomials of degree at most $N-1$.

In Eqs. (\ref{eq:step_RH_split}), we have managed to split the original equation, which combined two singular components with singularities at different points, into two equations, each containing only one of the sets of singularities.  Each of Eqs. (\ref{eq:step_RH_split}) can now be subjected to the regularisation procedure described in Part I.

%%%%%%%%%%%%%%%%%%%%%%%%%%%%

\subsection{Regularisation of the equation
\label{sub:Regularisation_of_the_equation}}

Following \S\S IV C-D of Part I, we first split the solutions $\psi^{\sigma}(k)$ into parts which are singular and regular on the real axis; the singular parts are generated by the propagating waves which are non-vanishing at infinity, while the regular parts are related to the evanescent waves which are only prominent near $x=0$.  We write:
\begin{equation}
\psi^{\sigma}(k) = \alpha^{\sigma}(k) + \sum_{j=1}^{N_{\sigma}^{r}} \mathcal{A}_{j}^{\sigma} \left[ \frac{1}{2} \delta \left(k-k_{L,j}^{r}\right) + \frac{s_{\sigma}}{2\pi \,i} \mathcal{P} \frac{1}{k-k_{\sigma,j}^{r}} \right] \,.
\label{eq:psi_singular_decomposition}
\end{equation}
The terms in square brackets are the half-Fourier transforms of plane waves, so that the coefficients $\mathcal{A}^{\sigma}_{j}$ are precisely the amplitudes of these plane waves\footnote{By ``plane waves'' here we simply mean the exponentials $e^{ikx}$ with unit amplitude.  They are not normalised; they become normalised upon multiplication by $\left|g_{\sigma}^{\prime}\left(k_{\sigma,j}^{r}\right)\right|^{-1/2}$.  Normalisation is taken into account, as in Part I, by a simple transformation of the scattering matrix (see Eq. (\ref{eq:scattering_matrix_normalized})).} in the waveform $\phi(x)$.  The additional contributions $\alpha^{\sigma}(k)$ are regular for all real $k$.  Substituting Eqs. (\ref{eq:psi_singular_decomposition}) into Eqs. (\ref{eq:step_RH_split}) and integrating over the $\delta$ functions and poles, we find a pair of equations for the $\alpha_{\sigma}(k)$:
\begin{multline}
\tilde{g}_{\sigma}(k) \, \alpha^{\sigma}(k) + \kappa^{\sigma}(k) \int_{-\infty}^{+\infty} \mathrm{d}k^{\prime} \, \alpha(k^{\prime}) \\
+ \frac{1}{2} \sum_{j=1}^{N_{\sigma}^{r}} \mathcal{A}_{j}^{\sigma} \left[ \frac{s_{\sigma}}{i\,\pi} \frac{\tilde{g}_{\sigma}(k)}{k-k_{\sigma,j}^{r}} + \kappa^{\sigma}(k)\right] + \frac{1}{2} \sum_{j=1}^{N_{-\sigma}^{r}} \mathcal{A}_{j}^{-\sigma} \kappa^{\sigma}(k) + s_{\sigma} \, \sum_{j=1}^{N} P_{j} \, p_{j}(k) \, = \, 0 \,,
\label{eq:step_alpha_int}
\end{multline}
where we have defined $\alpha(k)=\alpha^{L}(k)+\alpha^{R}(k)$, and where we have replaced $\phi_{0}$ with its explicit form as an integral (see Eq. (\ref{eq:phi0_defn})).

While Eqs. (\ref{eq:step_alpha_int}) have the same general form as Eqs. (\ref{eq:step_RH_split}) and remain non-invertible, we have defined $\alpha^{\sigma}(k)$ to be regular for all real $k$.  Thus, when $k$ approaches a zero of $\tilde{g}_{\sigma}(k)$, the first term of one of Eqs. (\ref{eq:step_alpha_int}) vanishes.  This yields two sets of regularity conditions, one for the real roots of $\tilde{g}_{L}(k)$, and another for the real roots of $\tilde{g}_{R}(k)$:
\begin{multline}
\kappa^{\sigma}\left(k_{\sigma,i}^{r}\right) \int_{-\infty}^{+\infty} \mathrm{d}k^{\prime} \, \alpha(k^{\prime}) + \frac{1}{2} \sum_{j=1}^{N_{\sigma}^{r}} \mathcal{A}_{j}^{\sigma} \left[ \frac{s_{\sigma}}{i \, \pi} \tilde{g}_{\sigma}^{\prime}\left(k_{\sigma,i}^{r}\right) \delta_{ij} + \kappa^{\sigma}\left(k_{\sigma,i}^{r}\right) \right] \\
+ \frac{1}{2} \sum_{j=1}^{N_{-\sigma}^{r}} \mathcal{A}_{j}^{-\sigma} \kappa^{\sigma}\left(k_{\sigma,i}^{r}\right) + s_{\sigma} \sum_{j=1}^{N} P_{j} \, p_{j}\left(k_{\sigma,i}^{r}\right) \, = \, 0 \,.
\label{eq:step_reg_conds}
\end{multline}
As in \S IV D of Part I (using a procedure inspired by Bart and Warnock \cite{Bart-Warnock-1973}), we note that Eqs. (\ref{eq:step_alpha_int}) can be transformed into invertible equations by subtraction of regularity conditions (\ref{eq:step_reg_conds}) times functions $f_{j}^{\sigma}(k)$ such that $f_{j}^{\sigma}\left(k_{\sigma,i}^{r}\right) = \delta_{ij}$, followed by division by $\tilde{g}_{\sigma}(k)$.  A convenient such set of functions can be constructed from the coefficient functions $\tilde{g}_{\sigma}(k)$ themselves:
\begin{equation}
f_{j}^{\sigma}(k) = \frac{\tilde{g}_{\sigma}(k)}{\left(k-k_{\sigma,j}^{r}\right) \, \tilde{g}^{\prime}_{\sigma}\left(k_{\sigma,j}^{r}\right)} \,.
\label{eq:unit_functions}
\end{equation}
Applying this procedure leads to the equations
\begin{multline}
\alpha^{\sigma}(k) + \bar{\kappa}^{\sigma}(k) \int_{-\infty}^{+\infty} \mathrm{d}k^{\prime} \, \alpha(k^{\prime}) \\
+ \frac{1}{2} \bar{\kappa}^{\sigma}(k) \left( \sum_{j=1}^{N_{\sigma}^{r}} \mathcal{A}_{j}^{\sigma} + \sum_{j=1}^{N_{-\sigma}^{r}} \mathcal{A}_{j}^{-\sigma} \right) + s_{\sigma} \sum_{j=1}^{N} P_{j} \, \bar{p}_{j}^{\sigma}(k) \, = \, 0 \,.
\label{eq:step_regular_alpha}
\end{multline}
Overbars -- in conjunction with $\sigma$ superscripts -- have been used to indicate that functions have been transformed according to the procedure outlined above: for an arbitrary function $F(k)$, we have
\begin{equation}
\bar{F}^{\sigma}(k) = \frac{F(k)}{\tilde{g}_{\sigma}(k)} - \sum_{j=1}^{N_{\sigma}^{r}} \frac{F\left(k_{\sigma,j}^{r}\right)}{\left(k-k_{\sigma,j}^{r}\right) \, \tilde{g}_{\sigma}^{\prime}\left(k_{\sigma,j}^{r}\right)} \,.
\label{eq:overbar_defn}
\end{equation}
Finally, we may add the two of Eqs. (\ref{eq:step_regular_alpha}) (with $\sigma=L$ and $\sigma=R$) together to get a single invertible equation for the single unknown function $\alpha(k)$:
\begin{equation}
\alpha(k) + \bar{\kappa}(k) \int_{-\infty}^{+\infty} \mathrm{d}k^{\prime} \, \alpha(k^{\prime}) + \frac{1}{2} \bar{\kappa}(k)\left(\sum_{j=1}^{N_{L}^{r}} \mathcal{A}_{j}^{L} + \sum_{j=1}^{N_{R}^{r}} \mathcal{A}_{j}^{R} \right) + \sum_{j=1}^{N} P_{j} \, \bar{p}_{j}(k) \, = \, 0 \,,
\label{eq:int_alpha_invertible}
\end{equation}
where we have defined
\begin{alignat}{2}
\bar{\kappa}(k) = \bar{\kappa}^{L}(k) + \bar{\kappa}^{R}(k) \,, & \qquad \bar{p}_{j}(k) = \bar{p}_{j}^{R}(k) - \bar{p}_{j}^{L}(k) \,.
\label{eq:overbarred_kappa_p_defns}
\end{alignat}
We note the minus sign that appears in the definition of $\bar{p}_{j}(k)$.  This can be understood by considering the degenerate case $u_{L}=u_{R}$: since in that case $\phi(x)$ is simply a sum of plane waves, $\alpha(k)$ must be zero; and similarly, the vanishing of $K_{\mathrm{step}}(k)$ in Eq. (\ref{eq:K_step}) means that $\bar{\kappa}(k)$ is zero.  This minus sign thus ensures that the last term on the left-hand side of Eq. (\ref{eq:int_alpha_invertible}) also vanishes, maintaining equality with the right-hand side.

%%%%%%%%%%%%%%%%%%%%%%%%%%%%

\subsection{Inversion of the integral operator
\label{sub:Inversion_of_the_integral_operator}}

Equation (\ref{eq:int_alpha_invertible}) can be written in the form
\begin{equation}
\int_{-\infty}^{+\infty} \mathrm{d}k^{\prime} \, \left[ \delta(k-k^{\prime}) + \bar{\kappa}(k) \right] \alpha(k^{\prime}) = -\sum_{j=1}^{N} P_{j}\,\bar{p}_{j}(k) - \frac{1}{2} \bar{\kappa}(k) \, \left( \sum_{j=1}^{N_{L}^{r}} \mathcal{A}_{j}^{L} + \sum_{j=1}^{N_{R}^{r}} \mathcal{A}_{j}^{R} \right) \,,
\label{eq:step_alpha_eqn}
\end{equation}
which is the continuous limit of a matrix equation relating the vector $\alpha\left(k_{m}\right)$ to $\bar{p}_{j}\left(k_{m}\right)$ and $\bar{\kappa}\left(k_{m}\right)$.  The kernel of the integral operator, $\delta(k-k^{\prime}) + \bar{\kappa}(k)$, is likewise the continuous limit of a matrix, and is of a simple enough form that its inverse can be written explicitly: defining, for an arbitrary integrable function $f(k)$, its integral
\begin{equation}
I_{f} = \int_{-\infty}^{+\infty} \mathrm{d}k \, f(k) \,,
\label{eq:integral_defn}
\end{equation}
we have
\begin{subequations}
\begin{equation}
\int_{-\infty}^{+\infty} \mathrm{d}k^{\prime\prime} \, V(k,k^{\prime\prime}) \, \left[ \delta(k^{\prime\prime}-k^{\prime}) + \bar{\kappa}(k^{\prime\prime}) \right] = \delta(k-k^{\prime})
\end{equation}
where
\begin{equation}
V(k,k^{\prime\prime}) = \delta(k-k^{\prime\prime}) - \frac{\bar{\kappa}(k)}{1+I_{\bar{\kappa}}} \,.
\end{equation}
\end{subequations}
Applying this inverse to Eq. (\ref{eq:step_alpha_eqn}), we find
\begin{equation}
\alpha(k) = -\sum_{j=1}^{N} P_{j} \, \bar{p}_{j}(k) + \frac{\bar{\kappa}(k)}{1+I_{\bar{\kappa}}} \left[ \sum_{j=1}^{N} P_{j} \, I_{\bar{p}_{j}} - \frac{1}{2} \sum_{j=1}^{N_{L}^{r}} \mathcal{A}_{j}^{L} - \frac{1}{2} \sum_{j=1}^{N_{R}^{r}} \mathcal{A}_{j}^{R} \right] \,.
\label{eq:step_alpha_soln}
\end{equation}
Since $\bar{\kappa}(k)$ and $\bar{p}_{j}(k)$ are simply sums over poles in the complex plane, the integrals $I_{\bar{\kappa}}$ and $I_{\bar{p}_{j}}$ can also be evaluated explicitly:
\begin{subequations}\begin{eqnarray}
I_{\bar{\kappa}} & = & i\pi \left[ \sum_{q=1}^{N_{R}^{c}} \frac{\kappa^{R}\left(k_{R,q}^{c}\right)}{\tilde{g}_{R}^{\prime}\left(k_{R,q}^{c}\right)} - \sum_{q=1}^{N_{L}^{c}} \frac{\kappa^{L}\left(k_{L,q}^{c}\right)}{\tilde{g}_{L}^{\prime}\left(k_{L,q}^{c}\right)} \right] \,, \\
I_{\bar{p}_{j}} & = & i\pi \left[ \sum_{q=1}^{N_{R}^{c}} \frac{p_{j}\left(k_{R,q}^{c}\right)}{\tilde{g}_{R}^{\prime}\left(k_{R,q}^{c}\right)} + \sum_{q=1}^{N_{L}^{c}} \frac{p_{j}\left(k_{L,q}^{c}\right)}{\tilde{g}_{L}^{\prime}\left(k_{L,q}^{c}\right)} \right] \,.
\end{eqnarray}\end{subequations}

To transform Eq. (\ref{eq:step_alpha_soln}) -- a solution for $\alpha(k)$ in terms of the coefficients of the propagating waves and of the basis polynomials -- into a linear relation between the coefficients of the propagating waves, we must enforce the regularity conditions (\ref{eq:step_reg_conds}), which now act as consistency relations.  There are $N_{L}+N_{R} = 2N$ of these -- exactly the number required to reduce the $3N$ degrees of freedom present in Eq. (\ref{eq:step_alpha_soln}) to the $N$ degrees of freedom of the space of solutions of the wave equation.  Integrating Eq. (\ref{eq:step_alpha_soln}), utilising the definitions (\ref{eq:integral_defn}), then substituting in Eqs. (\ref{eq:step_reg_conds}) yields a set of $2N$ linear equations ($N_{L}^{r}$ for $\sigma=L$ plus $N_{R}^{r}$ for $\sigma=R$) in $3N$ unknowns:
\begin{multline}
\frac{1}{2}\sum_{j=1}^{N_{\sigma}^{r}} \mathcal{A}_{j}^{\sigma} \left[ \frac{s_{\sigma}}{i\,\pi} \tilde{g}_{\sigma}^{\prime}\left(k_{\sigma,i}^{r}\right) \delta_{ij} + \frac{\kappa^{\sigma}\left(k_{\sigma,i}^{r}\right)}{1+I_{\bar{\kappa}}} \right] + \frac{1}{2} \sum_{j=1}^{N_{-\sigma}^{r}} \mathcal{A}_{j}^{-\sigma} \frac{\kappa^{\sigma}\left(k_{\sigma,i}^{r}\right)}{1+I_{\bar{\kappa}}} \\
+ \sum_{j=1}^{N} P_{j} \left[ s_{\sigma} \, p_{j}\left(k_{\sigma,i}^{r}\right) - \frac{I_{\bar{p}_{j}} \, \kappa^{\sigma}\left(k_{\sigma,i}^{r}\right)}{1+I_{\bar{\kappa}}} \right] \, = \, 0 \,.
\label{eq:step_linear_3N_unknowns}
\end{multline}
For each value of $\sigma$, this is a linear equation relating the $N^{r}_{L}$-dimensional vector $\vec{A}^{L}$, the $N^{r}_{R}$-dimensional vector $\vec{A}^{R}$ and the $N$-dimensional vector $\vec{P}$:
\begin{equation}
\mathcal{M}^{\sigma}_{\,\,\sigma} \, \vec{\mathcal{A}}^{\sigma} + \mathcal{M}^{\sigma}_{\,\,-\sigma} \, \vec{\mathcal{A}}^{-\sigma} + \mathcal{P}^{\sigma} \, \vec{P} = 0 \,,
\label{eq:step_matrix_eqn}
\end{equation}
where, making use of the abbreviations
\begin{alignat}{3}
\tilde{g}^{\prime}_{\sigma,i} = \tilde{g}_{\sigma}\left(k_{\sigma,i}^{r}\right) \,, & \qquad \kappa^{\sigma}_{i} = \kappa^{\sigma}\left(k_{\sigma,i}^{r}\right) \,, & \qquad p^{\sigma}_{ij} = p_{j} \left(k_{\sigma,i}^{r}\right) \,,
\label{eq:abbreviations}
\end{alignat}
the matrix elements are given by
\begin{subequations}\begin{eqnarray}
\left[ \mathcal{M}^{\sigma}_{\,\,\sigma} \right]_{ij} & = & \frac{1}{2} \left[ \frac{s_{\sigma}}{i \, \pi} \, \tilde{g}^{\prime}_{\sigma,i} \, \delta_{ij} + \frac{\kappa_{i}^{\sigma}}{1+I_{\bar{\kappa}}} \right] \,, \\
\left[ \mathcal{M}^{\sigma}_{\,\,-\sigma} \right]_{ij} & = & \frac{1}{2} \, \frac{\kappa_{i}^{\sigma}}{1+I_{\bar{\kappa}}} \,, \\
\left[ \mathcal{P}^{\sigma} \right]_{ij} & = & s_{\sigma} \, p_{ij}^{\sigma} - \frac{\kappa_{i}^{\sigma} \, I_{\bar{p}_{j}}}{1+I_{\bar{\kappa}}} \,.
\end{eqnarray}\label{eq:step_matrix_elements}\end{subequations}

%%%%%%%%%%%%%%%%%%%%%%%%%%%%

\subsection{Rearrangement into the scattering matrix
\label{sub:Rearrangement_into_the_scattering_matrix}}

Although we have managed to reduce the problem to a linear equation of the form (\ref{eq:step_matrix_eqn}), it is not yet in soluble form because in general the unknown vectors have different dimensions and the matrices are not square.  It is for this reason that the in- and out-bases are so useful, since they necessarily have the same dimension.  Therefore, we seek a way of rearranging the linear system (\ref{eq:step_matrix_eqn}) in terms of $\vec{\mathcal{A}}^{\mathrm{in}}$ and $\vec{\mathcal{A}}^{\mathrm{out}}$, making the matrices square and allowing inversion of them so that $\vec{P}$ can be eliminated.  To this end, we introduce the label $\rho$ which takes the values `in' and `out' to indicate whether a given wave is ingoing or outgoing, and relabel the waves and their amplitudes using $\rho$ rather than $\sigma$.  This relabelling can be described by a set of four (generally non-square) projection operators $Q^{\rho}_{\,\,\sigma}$: the $\left[ij\right]$ element of $Q^{\rho}_{\,\,\sigma}$ is equal to $1$ if the wavevector $k_{\rho,i}$ is the same as the wavevector $k_{\sigma,j}$, and $0$ otherwise.  The vectors $\vec{\mathcal{A}}^{\rho}$ are then related to the vectors $\vec{\mathcal{A}}^{\sigma}$ via
\begin{equation}
\vec{\mathcal{A}}^{\rho} = Q^{\rho}_{\,\,\sigma} \, \vec{\mathcal{A}}^{\sigma} + Q^{\rho}_{\,\,-\sigma} \, \vec{\mathcal{A}}^{-\sigma} \,.
\label{eq:sigma_to_rho}
\end{equation}
The inverse of this equation is effected by means of projection operators $Q^{\sigma}_{\,\,\rho}$, which are just the transposes of the projection operators in (\ref{eq:sigma_to_rho}):
\begin{alignat}{2}
\vec{\mathcal{A}}^{\sigma} = Q^{\sigma}_{\,\,\rho} \, \vec{\mathcal{A}}^{\rho} + Q^{\sigma}_{\,\,-\rho} \, \vec{\mathcal{A}}^{-\rho} \,, & \qquad Q^{\sigma}_{\,\,\rho} = \left[ Q^{\rho}_{\,\,\sigma} \right]^{T} \,.
\label{eq:rho_to_sigma}
\end{alignat}

The $\sigma$ component of Eq. (\ref{eq:step_matrix_eqn}) has dimension $N_{\sigma}^{r}$, and is compatible with left multiplication by $Q^{\rho}_{\,\,\sigma}$ for both values of $\rho$.  Since each row and column of $Q^{\rho}_{\,\,\sigma}$ has at most one non-zero element, this corresponds to selecting some of the rows of Eq. (\ref{eq:step_matrix_eqn}) and setting the others to zero.  However, since each wavevector lies in one of the asymptotic regions, each zero row of $Q^{\rho}_{\,\,\sigma}$ must be non-zero in $Q^{\rho}_{\,\,-\sigma}$, and each non-zero row of $Q^{\rho}_{\,\,\sigma}$ must be zero in $Q^{\rho}_{\,\,-\sigma}$.  Therefore, left multiplication of the $\sigma$ component of (\ref{eq:step_matrix_eqn}) by $Q^{\rho}_{\,\,\sigma}$ rearranges the equations into $N$ rows, some of which are zero; and left multiplication of the $-\sigma$ component of (\ref{eq:step_matrix_eqn}) by $Q^{\rho}_{\,\,-\sigma}$ does the same thing, with the zero rows and non-zero rows switched.  Adding the results together gives a rearranged set of $N$ rows of equations, none of which vanishes.  Performing the same manipulation with $\rho$ replaced by $-\rho$, we recover a linear system similar to (\ref{eq:step_matrix_eqn}), but with all vectors of dimension $N$ and all matrices $N \times N$:
\begin{equation}
\mathcal{M}^{\rho}_{\,\,\rho} \, \vec{\mathcal{A}}^{\rho} + \mathcal{M}^{\rho}_{\,\,-\rho} \, \vec{\mathcal{A}}^{-\rho} + \mathcal{P}^{\rho} \, \vec{P} = 0 \,,
\label{eq:step_matrix_in-out}
\end{equation}
where the matrices are related to those with $\sigma$ labels via
\begin{alignat}{2}
\mathcal{M}^{\rho}_{\,\,\rho^{\prime}} = \sum_{\sigma,\sigma^{\prime}} Q^{\rho}_{\,\,\sigma} \, \mathcal{M}^{\sigma}_{\,\,\sigma^{\prime}} \, Q^{\sigma^{\prime}}_{\,\,\rho^{\prime}} \,, & \qquad \mathcal{P}^{\rho} = \sum_{\sigma} Q^{\rho}_{\,\,\sigma} \, \mathcal{P}^{\sigma} \,.
\label{eq:matrix_sigma_to_rho}
\end{alignat}
The matrices $\mathcal{P}^{\rho}$ are invertible, and we can solve both of Eqs. (\ref{eq:step_matrix_in-out}) for $\vec{P}$.  Setting these equal, straightforward manipulation yields the sought-for equation
\begin{equation}
\vec{\mathcal{A}}^{\mathrm{out}} = \mathcal{S} \, \vec{\mathcal{A}}^{\mathrm{in}}
\label{eq:scattering_matrix_eqn}
\end{equation}
with the scattering matrix
\begin{equation}
\mathcal{S} = \left[ \mathcal{P}^{\mathrm{out}\,-1}\mathcal{M}^{\mathrm{out}}_{\,\,\mathrm{out}} - \mathcal{P}^{\mathrm{in}\,-1}\mathcal{M}^{\mathrm{in}}_{\,\,\mathrm{out}} \right]^{-1} \, \left[ \mathcal{P}^{\mathrm{in}\,-1}\mathcal{M}^{\mathrm{in}}_{\,\,\mathrm{in}} - \mathcal{P}^{\mathrm{out}\,-1}\mathcal{M}^{\mathrm{out}}_{\,\,\mathrm{in}} \right] \,.
\label{eq:step_scattering_matrix}
\end{equation}

Finally, if required, we can transform $\mathcal{S}$ into a scattering matrix $\mathcal{S}_{N}$ between normalized waves.  The normalization factor for the plane wave with wavevector $k_{\sigma,i}^{r}$ is $\mathcal{N}_{\sigma,i} = \left| g^{\prime}_{\sigma}\left(k_{\sigma,i}^{r}\right) \right|^{-1/2}$.  Arranging these into two diagonal matrices $\hat{\mathcal{N}}^{\mathrm{in}}$ and $\hat{\mathcal{N}}^{\mathrm{out}}$, we have
\begin{equation}
\mathcal{S}_{N} = \hat{\mathcal{N}}^{\mathrm{out}\,-1} \, \mathcal{S} \, \hat{\mathcal{N}}^{\mathrm{in}} \,.
\label{eq:scattering_matrix_normalized}
\end{equation}

%%%%%%%%%%%%%%%%%%%%%%%%%%%%

\subsection{Novelty of the integral method
\label{sub:Novelty_of_the_integral_method}}

Although numerics are required to evaluate the roots of the dispersion relation and to perform the matrix operations required to reach the final scattering matrix, the solution calculated here is in principal exact, limited only by numerical precision.  In particular, the integrals can be performed analytically; there is no need for discretisation to perform numerical integration or numerical inversion of an integral kernel, both of which have an accuracy limited by the spacing of the discretised grid.

It should be noted that the availability of an exact solution for the step discontinuous background is not new.  From an ODE in position space, one can derive matching conditions for $\phi_{\omega}(x)$ at the discontinuity, resulting in a system of linear equations relating the amplitudes of the plane waves on either side \cite{Robertson-2012,Corley-1997,Recati-et-al-2009,Finazzi-Parentani-2012}.  The method we have outlined here is equivalent, differing simply in that it views the problem from Fourier space rather than position space.  It turns out that this allows a neater way of dealing with the complex wavevectors, for while these must be explicitly included in the position space solution, resulting in a linear system of potentially very large dimension, the Fourier space approach wraps up their entire contribution in the $N+1$ integrals $I_{\bar{\kappa}}$ and $I_{\bar{p}_{j}}$ which appear in the matrix elements of Eqs. (\ref{eq:step_matrix_elements}).  It cannot be said, however, that this significantly improves the efficiency of the numerical procedure.  Rather, the Fourier space approach to the step discontinuous background should be viewed as a precursor to the Fourier space approach to an arbitrary background, to which we now turn.

%%%%%%%%%%%%%%%%%%%%%%%%%%%%%%%%%%%%%%%%%%%%%%%%%%%

\section{General background
\label{sec:General_background}}

Having solved for the scattering matrix of a step discontinuous background, and having introduced the additional machinery necessary to separate the singular operators associated with the left- and right-hand asymptotic regions, we now set about generalising the procedure to an arbitrary background which is asymptotically asymmetric.  Since much of the development here is directly analogous to that of Section \ref{sec:Step-discontinuous_background}, such points are treated here only briefly, while attention is paid to the notable differences.

%%%%%%%%%%%%%%%%%%%%%%%%%%%%

\subsection{Splitting into half-plane analytic parts
\label{sub:General_Splitting_into_half-plane_analytic_parts}}

We shall assume that the background is continuous at $x=0$, so that $K_{\mathrm{step}}(k)$ of Eq. (\ref{eq:kernel_decomposition}) vanishes.  Our starting point is therefore the first line of Eq. (\ref{eq:int_eqn_diff_asymptotics}).  The kernel $K(k,k^{\prime})$ can be split into two parts which are analytic and $\sim 1/k^{\prime}$ on opposite halves of the complex $k^{\prime}$-plane.  Division by the ``forbidden'' roots of the dispersion relations proceeds as before, but instead of a single function of a single variable in Eq. (\ref{eq:kappa_defn}), we now have two functions of two variables:
\begin{equation}
\kappa_{\sigma}(k,k^{\prime}) = \frac{K_{\sigma}(k,k^{\prime})}{\prod_{q=1}^{N_{L}^{c}}\left(k-k_{L,q}^{c\star}\right) \, \prod_{q=1}^{N_{R}^{c}} \left(k-k_{R,q}^{c\star}\right)} \,,
\label{eq:kappaLR_defn}
\end{equation}
where, as before, $\sigma$ takes the values $L$ and $R$, and as a subscript indicates the analyticity in the primed argument $k^{\prime}$.  Following the procedure outlined in \S\ref{sub:Splitting_into_half-plane_analytic_parts}, we find that we have to split $\kappa_{\sigma}(k,k^{\prime})$ into half-plane analytic parts of the {\it un}primed variable $k$.  Although they do not behave like $1/k$ asymptotically, the $K_{\sigma}(k,k^{\prime})$ are analytic on a half-plane, so what remains is to rearrange the poles at the ``forbidden'' wavevectors.  We can write
\begin{equation}
\frac{1}{\prod_{q=1}^{N_{L}^{c}}\left(k-k_{L,q}^{c\star}\right) \, \prod_{q=1}^{N_{R}^{c}} \left(k-k_{R,q}^{c\star}\right)} = \sum_{q=1}^{N_{L}^{c}} \frac{C_{L,q}^{-1}}{k-k_{L,q}^{c\star}} + \sum_{q=1}^{N_{R}^{c}} \frac{C_{R,q}^{-1}}{k-k_{R,q}^{c\star}} \,,
\label{eq:poles_separated}
\end{equation}
where the coefficients $C_{\sigma,j}$ are given in Eqs. (\ref{eq:kappa_split}).  Of the terms of the right-hand side of Eq. (\ref{eq:poles_separated}), the first is analytic on the lower half plane, and the second on the upper half plane.  On multiplication by $K_{\sigma}(k,k^{\prime})$, one of the two products is automatically analytic on a half plane, while the other is non-analytic only due to a discrete set of poles, which can be subtracted and moved over to the other term.  Explicitly, we have:
\begin{equation}
\kappa_{\sigma}(k,k^{\prime}) = \kappa^{\sigma}_{\,\,\sigma}(k,k^{\prime}) + \kappa^{-\sigma}_{\,\,\sigma}(k,k^{\prime}) \,,
\label{eq:general_kappa_split}
\end{equation}
where
\begin{subequations}\begin{eqnarray}
\kappa^{\sigma}_{\,\,\sigma}(k,k^{\prime}) & = & \sum_{q=1}^{N_{-\sigma}^{c}} C_{-\sigma,q}^{-1} \, \frac{K_{\sigma}(k,k^{\prime})}{k-k_{-\sigma,q}^{c\star}} + \sum_{q=1}^{N_{\sigma}^{c}} C_{\sigma,q}^{-1} \, \frac{K_{\sigma}(k,k^{\prime}) - K_{\sigma}\left(k_{\sigma,q}^{c\star},k^{\prime}\right)}{k-k_{\sigma,q}^{c\star}} \,, \label{eq:general_kappa_equal_sigma} \\
\kappa^{-\sigma}_{\,\,\sigma}(k,k^{\prime}) & = & \sum_{q=1}^{N_{\sigma}^{c}} C_{\sigma,q}^{-1} \, \frac{K_{\sigma}\left(k_{\sigma,q}^{c\star},k^{\prime}\right)}{k-k_{\sigma,q}^{c\star}} \,. \label{eq:general_kappa_different_sigma}
\end{eqnarray}\label{eq:general_kappa_defns}\end{subequations}
Note that we have assumed that we know not only the boundary value of $K_{\sigma}(k,k^{\prime})$ when $k$ is real, but also its values for complex $k$ on the half plane where it is analytic.

To be clear about notation, we emphasise here that the $\sigma$ {\it sub}script refers to which half-Fourier transform of the waveform the particular kernel acts on, and thus to its analyticity in the primed coordinate $k^{\prime}$ ({\it i.e.} that which is integrated over).  As a {\it super}script, however, $\sigma$ refers to the analyticity in the {\it un}primed coordinate, and points to the corresponding half-Fourier transform of the waveform which has the same analyticity.

Splitting the kernels according to Eqs. (\ref{eq:general_kappa_split})-(\ref{eq:general_kappa_defns}), we are once again led to an equality between two functions of $k$ which are manifestly analytic on opposite half planes, and are thus equal to an entire function.  By considering the asymptotic behaviour of this function, we are led to the conclusion that it must be a polynomial of degree at most $N-1$, where $N$ is the number of ingoing or outgoing waves.  We can thus write the generalised versions of Eqs. (\ref{eq:step_RH_split}):
\begin{subequations}\begin{multline}
\tilde{g}_{L}(k) \psi^{L}(k) + \int_{-\infty}^{+\infty} \mathrm{d}k^{\prime} \, \kappa_{\,\,L}^{L}(k,k^{\prime}) \psi^{L}(k^{\prime}) + \int_{-\infty}^{+\infty} \mathrm{d}k^{\prime} \, \kappa_{\,\,R}^{L}(k,k^{\prime}) \psi^{R}(k^{\prime})  \\
=  \sum_{j=1}^{N} P_{j} \, p_{j}(k) \,,
\end{multline}
\begin{multline}
\tilde{g}_{R}(k) \psi^{R}(k) + \int_{-\infty}^{+\infty} \mathrm{d}k^{\prime} \, \kappa_{\,\,R}^{R}(k,k^{\prime}) \psi^{R}(k^{\prime}) + \int_{-\infty}^{+\infty} \mathrm{d}k^{\prime} \, \kappa_{\,\,L}^{R}(k,k^{\prime}) \psi^{L}(k^{\prime}) \\
=  -\sum_{j=1}^{N} P_{j} \, p_{j}(k) \,.
\end{multline}
\label{eq:general_split_int_eqns}\end{subequations}

%%%%%%%%%%%%%%%%%%%%%%%%%%%%

\subsection{Regularisation of the equation
\label{sub:General_Regularisation_of_the_equation}}

Upon splitting the solutions $\psi^{\sigma}(k)$ into their singular and regular parts as in Eq. (\ref{eq:psi_singular_decomposition}) and substituting into Eqs. (\ref{eq:general_split_int_eqns}), we are led to the analogue of Eqs. (\ref{eq:step_alpha_int}):
\begin{multline}
\tilde{g}_{\sigma}(k) \, \alpha^{\sigma}(k) + \int_{-\infty}^{+\infty} \mathrm{d} k^{\prime} \, \kappa^{\sigma}(k,k^{\prime}) \, \alpha(k^{\prime}) + s_{\sigma} \sum_{j=1}^{N} P_{j} \, p_{j}(k) \\
+ \sum_{j=1}^{N_{\sigma}^{r}} \mathcal{A}_{j}^{\sigma} \left[ \frac{s_{\sigma}}{2\pi i} \, \frac{\tilde{g}_{\sigma}(k)}{k-k_{\sigma,j}^{r}} + \kappa_{\,\,\sigma}^{\sigma}\left(k,k_{\sigma,j}^{r}\right) \right] + \sum_{j=1}^{N_{-\sigma}^{r}} \mathcal{A}_{j}^{-\sigma} \, \kappa_{\,\,-\sigma}^{\sigma}\left(k,k_{-\sigma,j}^{r}\right) \, = \, 0 \, .
\label{eq:general_alpha_int}
\end{multline}
Here, we have defined
\begin{equation}
\kappa^{\sigma}(k,k^{\prime}) = \sum_{\sigma^{\prime}} \kappa^{\sigma}_{\,\,\sigma^{\prime}}(k,k^{\prime})
\end{equation}
and noted that, due to their analyticity properties, the $\kappa^{\sigma}_{\,\,\sigma^{\prime}}(k,k^{\prime})$ and $\alpha^{\sigma}(k^{\prime})$ components of the integrand can be added separately because the cross terms vanish upon integration.  By definition, the $\alpha^{\sigma}(k)$ are regular on the real axis, so sending $k \rightarrow k_{\sigma,i}^{r}$ yields two sets of regularity conditions, $N_{L}^{r}$ for $\sigma=L$ and $N_{R}^{r}$ for $\sigma=R$:
\begin{multline}
\int_{-\infty}^{+\infty} \mathrm{d}k^{\prime} \, \kappa^{\sigma}\left(k_{\sigma,i}^{r},k^{\prime}\right) \, \alpha(k^{\prime}) + s_{\sigma} \, \sum_{j=1}P_{j}\,p_{j}\left(k_{\sigma,i}^{r}\right) \\
+ \sum_{j=1}^{N_{\sigma}^{r}} \mathcal{A}_{j}^{\sigma} \left[ \frac{s_{\sigma}}{2\pi i} \, \tilde{g}_{\sigma}^{\prime}\left(k_{\sigma,i}^{r}\right) \delta_{ij} + \kappa_{\,\,\sigma}^{\sigma}\left(k_{\sigma,i}^{r},k_{\sigma,j}^{r}\right)\right] + \sum_{j=1}^{N_{-\sigma}^{r}} \mathcal{A}_{j}^{-\sigma} \, \kappa_{\,\,-\sigma}^{\sigma}\left(k_{\sigma,i}^{r},k_{-\sigma,j}^{r}\right) = 0 \,.
\label{eq:general_reg_conds}
\end{multline}
By the same procedure used in \S\ref{sub:Regularisation_of_the_equation} -- whereby each of the regularity conditions (\ref{eq:general_reg_conds}) is multiplied by the corresponding function (\ref{eq:unit_functions}) and subtracted from Eq. (\ref{eq:general_alpha_int}), after which $\tilde{g}_{\sigma}(k)$ can be divided out -- we are led to the analogue of Eqs. (\ref{eq:step_regular_alpha}):
\begin{multline}
\alpha^{\sigma}(k) + \int_{-\infty}^{+\infty} \mathrm{d}k^{\prime} \, \bar{\kappa}^{\sigma}(k,k^{\prime}) \, \alpha(k^{\prime}) \\
+ \sum_{j=1}^{N_{\sigma}^{r}} \mathcal{A}_{j}^{\sigma} \, \bar{\kappa}_{\,\,\sigma}^{\sigma}\left(k,k_{\sigma,j}^{r}\right) + \sum_{j=1}^{N_{-\sigma}^{r}} \mathcal{A}_{j}^{-\sigma} \, \bar{\kappa}_{\,\,-\sigma}^{\sigma}\left(k,k^{r}_{-\sigma,j}\right) + s_{\sigma} \sum_{j=1}^{N} P_{j} \, \bar{p}^{\sigma}_{j}(k) = 0 \,,
\label{eq:general_regular_alpha}
\end{multline}
where the definition of the overbar is the same as in Eq. (\ref{eq:overbar_defn}).  As in \S\ref{sub:Regularisation_of_the_equation}, we can add the $\sigma=L$ and $\sigma=R$ components of Eq. (\ref{eq:general_regular_alpha}) together, yielding a single equation for $\alpha(k)=\alpha^{L}(k)+\alpha^{R}(k)$:
\begin{multline}
\alpha(k) + \int_{-\infty}^{+\infty} \mathrm{d}k^{\prime} \, \bar{\kappa}(k,k^{\prime}) \, \alpha(k^{\prime}) \\
+ \sum_{j=1}^{N_{L}^{r}} \mathcal{A}_{j}^{L} \, \bar{\kappa}_{L}\left(k,k_{L,j}^{r}\right) \, + \, \sum_{j=1}^{N_{R}^{r}} \mathcal{A}_{j}^{R} \, \bar{\kappa}_{R}\left(k,k_{R,j}^{r}\right) + \sum_{j=1}^{N} P_{j} \, \bar{p}_{j}(k) = 0 \,,
\label{eq:general_int_alpha_invertible}
\end{multline}
where we have defined
\begin{alignat}{2}
\bar{\kappa}_{\sigma}(k,k^{\prime}) = \sum_{\sigma^{\prime}} \bar{\kappa}^{\sigma^{\prime}}_{\,\,\sigma}(k,k^{\prime}) \,, & \qquad \bar{\kappa}(k,k^{\prime}) = \sum_{\sigma} \sum_{\sigma^{\prime}} \bar{\kappa}^{\sigma^{\prime}}_{\,\,\sigma}(k,k^{\prime}) \,,
\end{alignat}
and where $\bar{p}_{j}(k)$ is defined as in Eqs. (\ref{eq:overbarred_kappa_p_defns}).

%%%%%%%%%%%%%%%%%%%%%%%%%%%%

\subsection{Inversion of the integral operator
\label{sub:General_Inversion_of_the_integral_operator}}

Equation (\ref{eq:general_int_alpha_invertible}) is invertible, {\it i.e.} for any set of coefficients, $\alpha(k)$ can be solved for uniquely.  However, given the general nature of the integral kernel $\bar{\kappa}(k,k^{\prime})$, we cannot invert the equation explicitly, and must instead resort to a numerical solution.  In this sense, Eq. (\ref{eq:general_int_alpha_invertible}) is more to be compared with Eq. (24) of Part I.  As there, we note that there exists an exact inverse kernel $V(k,k^{\prime})$ such that
\begin{equation}
\alpha(k) = -\int_{-\infty}^{+\infty} \mathrm{d}k^{\prime} \, V(k,k^{\prime}) \left[ \sum_{j=1}^{N_{L}^{r}} \mathcal{A}_{j}^{L} \, \bar{\kappa}_{L}\left(k^{\prime},k_{L,j}^{r}\right) + \sum_{j=1}^{N_{R}^{r}} \mathcal{A}_{j}^{R} \, \bar{\kappa}_{R}\left(k^{\prime},k_{R,j}^{r}\right) + \sum_{j=1}^{N} P_{j} \, \bar{p}_{j}(k^{\prime}) \right] \,.
\label{eq:general_alpha_soln_exact}
\end{equation}
Upon discretisation for the purposes of numerics, Eq. (\ref{eq:general_alpha_soln_exact}) becomes a matrix equation, the discretised kernels becoming matrices, and the discretised inverse kernel being
\begin{equation}
V = \left[ \delta_{nm} + \bar{\kappa}\left(k_{n},k_{m}\right) \cdot \Delta k_{m} \right]^{-1} \,.
\label{eq:discrete_inverse_kernel}
\end{equation}
This can be calculated efficiently using standard numerical algorithms.  The only difference from \S IV E of Part I is that we have allowed the distance between points $k_{m}$ on the integration grid to depend on $m$.  This is in anticipation of a change of variables to improve convergence of the integral (see Appendix \ref{app:Change_of_variables_for_numerical_integration}).

The regularity conditions (\ref{eq:general_reg_conds}), which at this point act as consistency relations, must now be enforced by substituting into them the solution (\ref{eq:general_alpha_soln_exact}).  This yields a set of $2N$ linear equations in the $3N$ unknowns $\mathcal{A}^{L}_{j}$, $\mathcal{A}^{R}_{j}$ and $P_{j}$.  To avoid cumbersome expressions, we adopt the abbreviations $\tilde{g}^{\prime}_{\sigma,i}$ and $p^{\sigma}_{ij}$ from Eqs. (\ref{eq:abbreviations}), as well as the following:
\begin{alignat}{3}
\kappa^{\sigma}_{\,\,\sigma^{\prime},ij} = \kappa^{\sigma}_{\,\,\sigma^{\prime}}\left(k_{\sigma,i}^{r},k_{\sigma^{\prime},j}^{r}\right) \,, & \qquad \kappa^{\sigma}_{i}(k) = \kappa^{\sigma}\left(k_{\sigma,i}^{r},k\right) \,, & \qquad \bar{\kappa}_{\sigma,j}(k) = \bar{\kappa}_{\sigma}\left(k,k_{\sigma,j}^{r}\right) \,.
\label{eq:abbreviations2}
\end{alignat}
The linear system of equations can now be written very succinctly as
\begin{equation}
\mathcal{M}^{\sigma}_{\,\,\sigma} \, \vec{A}^{\sigma} + \mathcal{M}^{\sigma}_{\,\,-\sigma} \, \vec{A}^{-\sigma} + \mathcal{P}^{\sigma} \, \vec{P} = 0 \,,
\label{eq:LR_matrix_equation}
\end{equation}
where the matrix elements are given by
\begin{subequations}\begin{eqnarray}
\left[ \mathcal{M}^{\sigma}_{\,\,\sigma} \right]_{ij} & = & \frac{s_{\sigma}}{2\pi i} \, \tilde{g}^{\prime}_{\sigma,i} \, \delta_{ij} + \kappa^{\sigma}_{\,\,\sigma,ij} - \int_{-\infty}^{+\infty} \int_{-\infty}^{+\infty} \mathrm{d}k \, \mathrm{d}k^{\prime} \, \kappa^{\sigma}_{i}(k) \, V(k,k^{\prime}) \, \bar{\kappa}_{\sigma,j}(k^{\prime}) \,, \qquad \\
\left[ \mathcal{M}^{\sigma}_{\,\,-\sigma} \right]_{ij} & = & \kappa^{\sigma}_{\,\,-\sigma,ij} - \int_{-\infty}^{+\infty} \int_{-\infty}^{+\infty} \mathrm{d}k \, \mathrm{d}k^{\prime} \, \kappa^{\sigma}_{i}(k) \, V(k,k^{\prime}) \, \bar{\kappa}_{-\sigma,j}(k^{\prime}) \,, \\
\left[ \mathcal{P}^{\sigma} \right]_{ij} & = & s_{\sigma} \, p^{\sigma}_{ij} - \int_{-\infty}^{+\infty} \int_{-\infty}^{+\infty} \mathrm{d}k \, \mathrm{d}k^{\prime} \, \kappa^{\sigma}_{i}(k) \, V(k,k^{\prime}) \, \bar{p}_{j}(k^{\prime}) \,.
\end{eqnarray}\label{eq:matrix_elements_exact}\end{subequations}
In discretised form, the integrals in Eqs. (\ref{eq:matrix_elements_exact}) become discrete sums, or equivalently products of matrices:
\begin{subequations}\begin{eqnarray}
\left[ \mathcal{M}^{\sigma}_{\,\,\sigma} \right]_{ij} & = & \frac{s_{\sigma}}{2\pi i} \, \tilde{g}^{\prime}_{\sigma,i} \, \delta_{ij} + \kappa^{\sigma}_{\,\,\sigma,ij} - \sum_{n,m} \, \Delta k_{n} \cdot \kappa^{\sigma}_{i}\left(k_{n}\right) \, V_{nm} \,\, \bar{\kappa}_{\sigma,j}\left(k_{m}\right) \,, \\
\left[ \mathcal{M}^{\sigma}_{\,\,-\sigma} \right]_{ij} & = & \kappa^{\sigma}_{\,\,-\sigma,ij} - \sum_{n,m} \, \Delta k_{n} \cdot \kappa^{\sigma}_{i}\left(k_{n}\right) \, V_{nm} \,\, \bar{\kappa}_{-\sigma,j}\left(k_{m}\right) \,, \\
\left[ \mathcal{P}^{\sigma} \right]_{ij} & = & s_{\sigma} \, p^{\sigma}_{ij} - \sum_{n,m} \, \Delta k_{n} \cdot \kappa^{\sigma}_{i}\left(k_{n}\right) \, V_{nm} \,\, \bar{p}_{j}\left(k_{m}\right) \,,
\end{eqnarray}\label{eq:matrix_elements_discretised}\end{subequations}
where, as in Part I, we have used the labels $m$ and $n$ to refer to points on the discretised integration grid, while $i$ and $j$ have been used to label the solutions of the dispersion relation.

Rearrangement of Eq. (\ref{eq:LR_matrix_equation}) in terms of the in- and out-bases proceeds exactly as in \S\ref{sub:Rearrangement_into_the_scattering_matrix}, leading again to the sought-for equations (\ref{eq:scattering_matrix_eqn})-(\ref{eq:scattering_matrix_normalized}).

%%%%%%%%%%%%%%%%%%%%%%%%%%%%

\subsection{Convergence of numerical integration
\label{sub:General_Convergenece_of_numerical_integration}}

In the general case considered here, the integrals that appear in Eqs. (\ref{eq:matrix_elements_exact}) converge slowly, because the various factors in the integrands typically behave asymptotically like $1/k$ or $1/k^{\prime}$.  Using a discretised integration with uniform spacing $\Delta k$ that is constant throughout the grid, this slow convergence would force us to take a large integration interval, and hence many points, greatly slowing down the numerical calculations.  Using such a large number of points to represent a simple $1/k$ behaviour would be very wasteful, and we seek a more economical approach.  One way would be, as in Part I, to cut off the discretised integration grid at a finite value of $k$ beyond which the integrands are approximately proportional to $1/k^{2}$, and to perform the remainder of the integral analytically.  This approach, however, complicates the numerical calculation of the inverse kernel $V(k,k^{\prime})$.

Instead, a change of variables is implemented (see Appendix \ref{app:Change_of_variables_for_numerical_integration} for details) which maps the infinite integration range of $k$ to a finite interval of a new variable $\zeta$.  Careful placing of the Jacobian of the variable transformation cancels out the $1/k$ convergence, so that the integrals to be evaluated can be considered as integrals of bounded functions over a finite interval.  These transformed integrals can then be evaluated numerically.  The spacing $\Delta \zeta$ is chosen to be uniform, inducing a non-uniform spacing $\Delta k_{n}$ in the $k$-representation.  Since the new integration range is finite, we do not need to choose the limits (as we did in Part I), but instead there is a parameter $k_{0}$ in the change of variables which determines the scale of the $\zeta$-representation near $k=0$.  The scaling parameter $k_{0}$ should be chosen such that (roughly speaking) the non-trivial part of the integral lies inside $(-k_{0},+k_{0})$, for the integrand outside this region is packed into a narrow region of $\zeta$-space, and should be as near as possible to the asymptotic $1/k$ behaviour in order to avoid sharp features whose resolution would require a very small $\Delta\zeta$.

%%%%%%%%%%%%%%%%%%%%%%%%%%%%%%%%%%%%%%%%%%%%%%%%%%%

\section{Application
\label{sec:Application}}

Here we shall illustrate the applicability of the methods described in \S\S \ref{sec:Step-discontinuous_background} and \ref{sec:General_background} by considering simple concrete examples.  These shall be based on the examples used in \S V of Part I, so that the results may be compared with those presented there.

%%%%%%%%%%%%%%%%%%%%%%%

\subsection{Dispersion relation
\label{sub:Dispersion_relation}}

As in Part I, we shall consider two dispersion relations: one is a low-degree polynomial which can be treated using standard ODE solution methods; the other is a more complicated function not amenable to such techniques, which we shall approximate within the relevant region of $k$-space as a polynomial of relatively high degree.  After suitable normalisation (see \S 5.1 of Part I), these dispersion relations are
\begin{alignat}{3}
c^{2}(k) = \frac{\mathrm{tanh}(k)}{k} & \qquad \mathrm{and} & \qquad c^{2}(k) = 1 - \frac{1}{3} k^{2} \,.
\label{eq:model_dispersion_relations}
\end{alignat}
The first describes surface waves in water when surface tension is negligible and the height of the water can be treated as constant \cite{Landau-Lifshitz-FM}; the second is chosen to be the second-order Taylor expansion of the first around $k=0$.

\begin{figure}
\includegraphics[width=0.8\columnwidth]{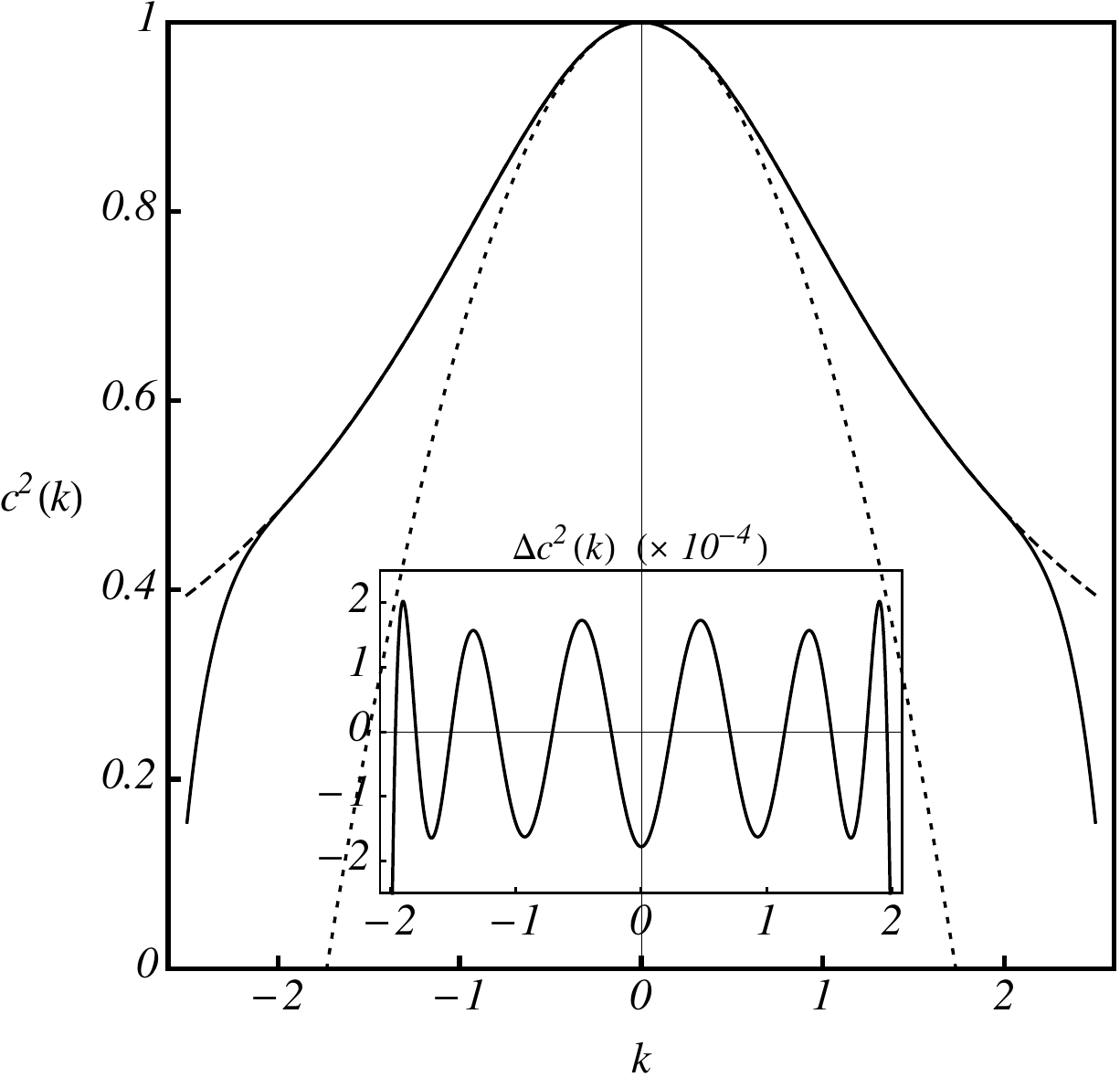}
\caption{\textsc{Dispersion relations}: \, Here are plotted the model dispersion relations of Eqs. (\ref{eq:model_dispersion_relations}), the first as a solid line and the second as a dotted line; note that these agree for small values of $k$.  The dashed line shows a polynomial approximation to the solid curve within the interval $k \in (-2,2)$, which entirely contains the region of the spectrum where Hawking radiation is predicted to occur (compare with Fig. \ref{fig:dispersion_in_stationary_frame}).  The approximation was found by discretising this interval into $200$ points, then performing a least-squares fit to a polynomial of degree $10$ with only even powers of $k$.  In the inset is plotted the difference between the polynomial approximation and the model dispersion relation we started with; in the region of interest, the approximation is found to agree with the model to about one part in $10^{4}$.
\label{fig:dispersion_relations}}
\end{figure}

On plotting the Doppler shifted dispersion relations as viewed in the stationary frame, the first of Eqs. (\ref{eq:model_dispersion_relations}) is found to predict a Hawking spectrum which is entirely contained within the interval $k \in (-2,2)$.  A polynomial approximation to this dispersion curve is found by discretising this interval into $200$ points and performing a least-squares fit of a polynomial of $10^{\mathrm{th}}$ degree.  The dispersion relations (\ref{eq:model_dispersion_relations}) and the polynomial approximation to the first are all plotted in Figure \ref{fig:dispersion_relations}.

%%%%%%%%%%%%%%%%%%%%%%%

\subsection{Velocity profile}

\begin{figure}
\includegraphics[width=0.8\columnwidth]{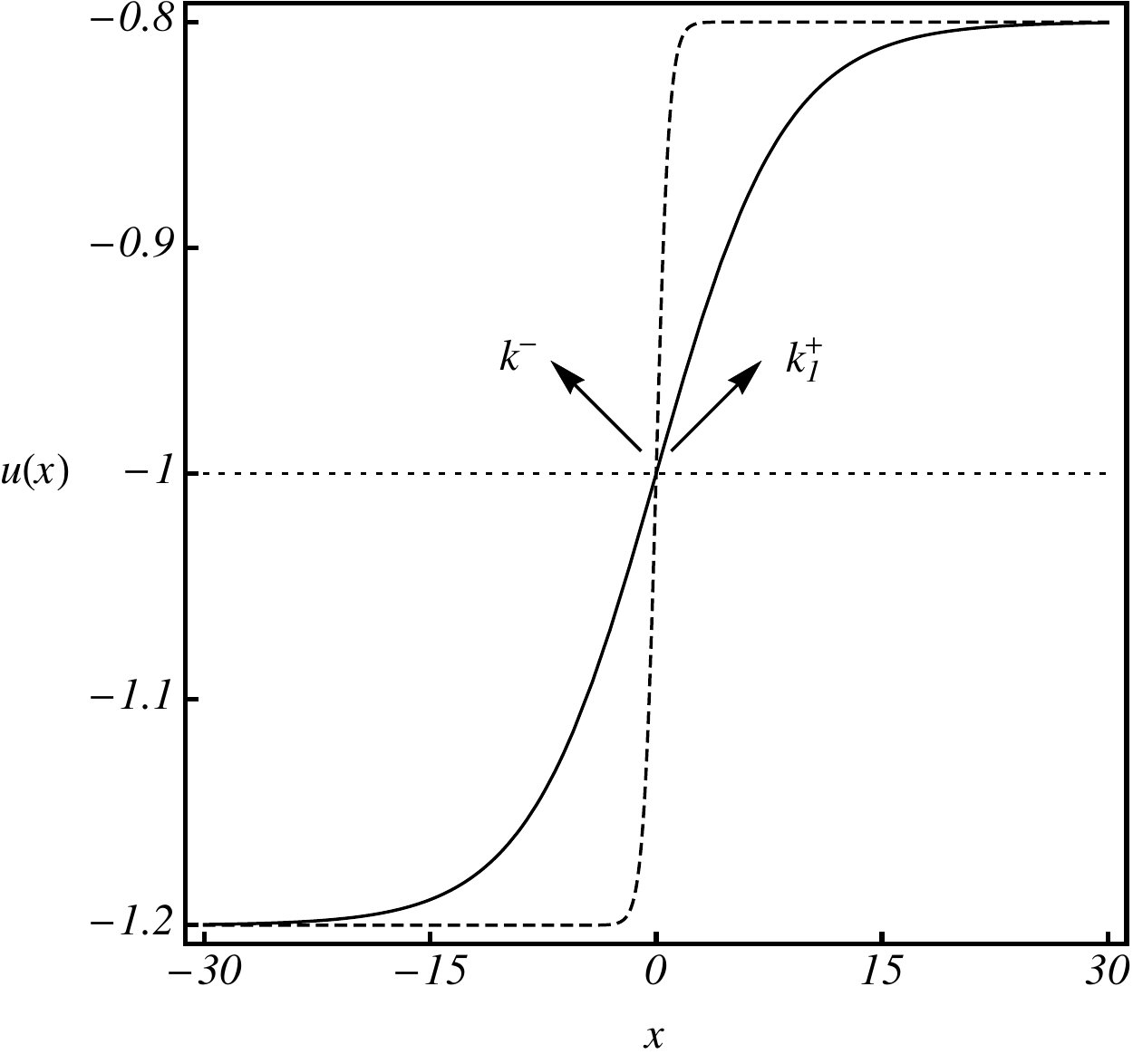}
\caption{\textsc{Velocity profiles}: \, Shown here are the flow velocity profiles of Eq. (\ref{eq:model_velocity_profile}), with $u_{R}=-0.8$, $u_{L}=-1.2$, and $a=0.118$ for the solid line and $a=1.18$ for the dashed line.  The dotted line plots $u=-1$, marking the transition between subsonic and supersonic flow; the point at which $u(x)$ crosses this is the event horizon.  It is a black hole horizon because the flow is accelerating in the direction of flow, and with it is associated spontaneous emission of Hawking pairs, the members of each pair being long-wavelength and emitted in opposite directions, as indicated here.  (See Fig. \ref{fig:dispersion_in_stationary_frame} for the positions of $k_{1}^{+}$ and $k^{-}$ in the dispersion relation.)
\label{fig:velocity_profiles}}
\end{figure}

The following form of the velocity profile is used, as it smoothly interpolates between two different asymptotic values:
\begin{equation}
u(x) = \frac{1}{2} \left( u_{R} + u_{L} \right) + \frac{1}{2} \left( u_{R} - u_{L} \right) \mathrm{tanh}\left( a x \right) \,.
\label{eq:model_velocity_profile}
\end{equation}
At $k=0$, the (normalised) wave speed is $1$, so where $|u|$ crosses $1$ is the analogue of the event horizon.  We take $u_{R} = -0.8$ and $u_{L} = -1.2$.  The medium is thus flowing to the left, and its speed increases in the direction of flow, so where $u=-1$ (at $x=0$) is a black hole horizon.  (There is no white hole horizon as in Part I: allowing for different asymptotic velocities allows us to study a single horizon in isolation.)  With $u_{R}$ and $u_{L}$ fixed, the remaining parameter $a$ controls the steepness of the profile -- or, by analogy with gravitational black holes, the ``surface gravity''.  As in Part I, we shall consider two values of $a$, one of which gives a slow variation of $u$ and the other a rapid variation.  For the sake of comparison, we shall choose these values such that the radiation temperatures given by the analogue of Hawking's original prediction \cite{Robertson-2012},
\begin{equation}
T=\frac{u^{\prime}\left(x_{h}\right)}{2\pi}
\label{eq:Hawking_prediction}
\end{equation}
where $x_{h}=0$ is the position of the horizon, are the same here as for the corresponding cases in Part I.  These predicted temperatures are $T=0.00375$ and $T=0.0375$ for the slowly-varying and rapidly-varying profiles, respectively, and correspond here to $a=0.118$ and $a=1.18$.  The velocity profiles are plotted in Figure \ref{fig:velocity_profiles}.

%%%%%%%%%%%%%%%%%%%%%%%

\subsection{Stationary-frame dispersion}

\begin{figure}
\subfloat{\includegraphics[width=0.45\columnwidth]{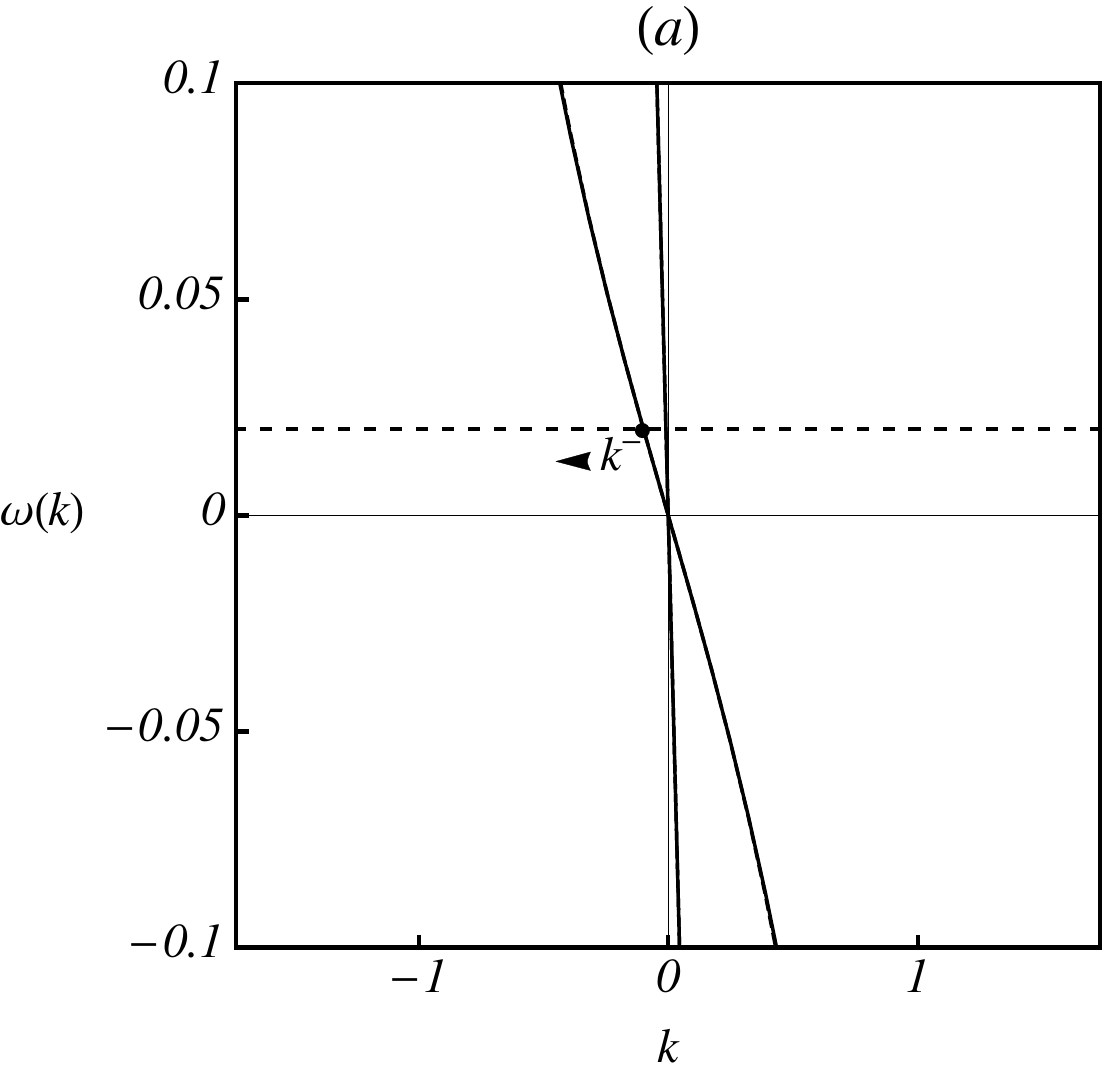} \, \includegraphics[width=0.45\columnwidth]{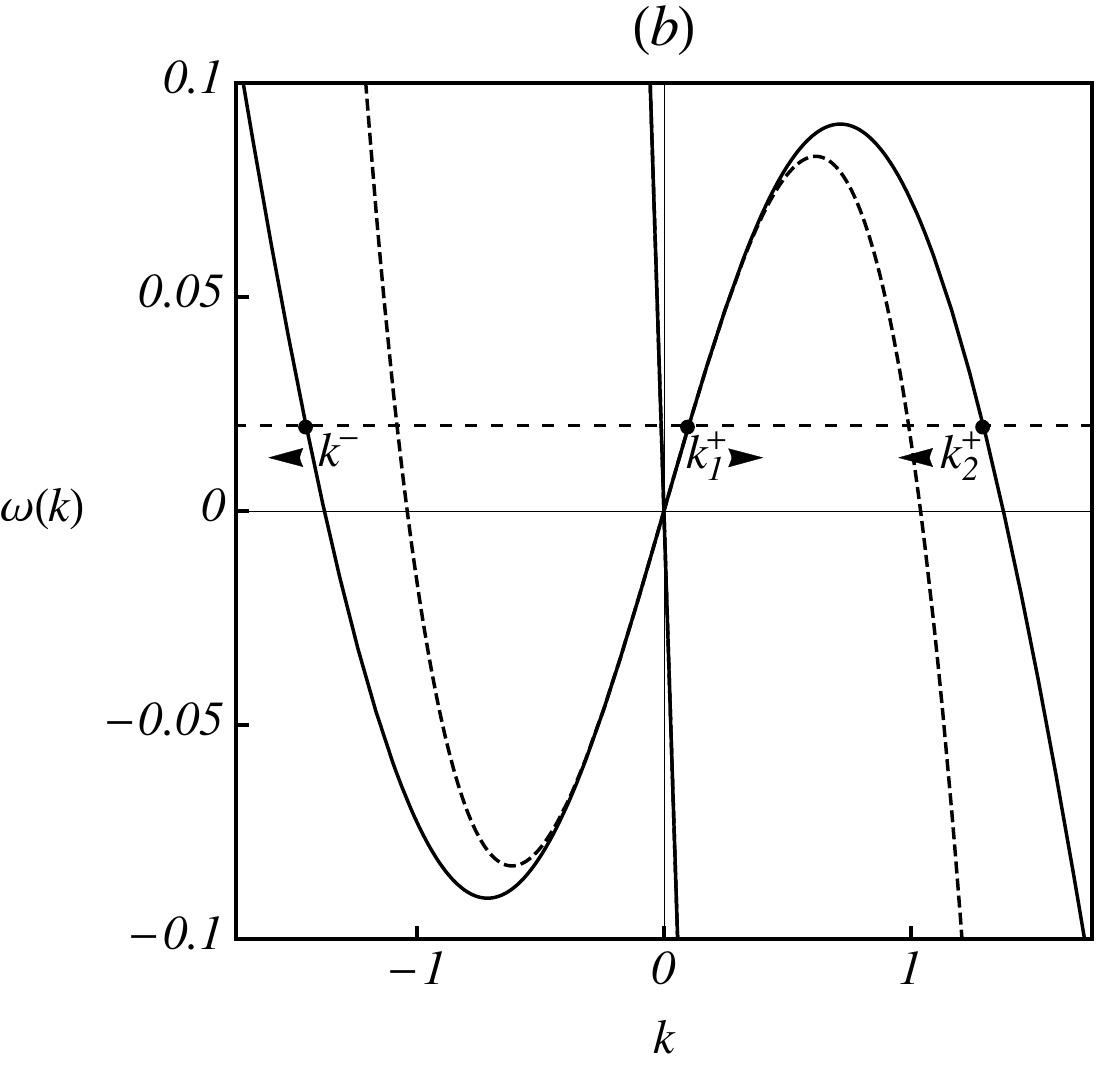}}
\caption{\textsc{Dispersion curves in stationary frame}: \, Here are plotted the Doppler shifted dispersion relations (\ref{eq:Doppler}) as viewed in the stationary frame, with $u=u_{L}=-1.2$ in $(a)$ and $u=u_{R}=-0.8$ in $(b)$.  The solid and dashed curves correspond to the first and second, respectively, of dispersion relations (\ref{eq:model_dispersion_relations}).  For a given conserved frequency $\omega$, there are several wavevector solutions, and these can be scattered into each other by an inhomogeneous flow.  We see here that, for $\omega$ less than some maximum value ($\sim 0.08$), there are two solutions on the left and three on the right; excluding the co-propagating waves, we have one on the left and three on the right, which are the solutions indicated here.  The group velocities of these solutions are the derivatives of the dispersion curves, so we can distinguish their directions; these are indicated here by small arrows.  Two of these wavevector solutions have group velocities pointing away from the scattering region around $x=0$, and are thus outgoing: $k^{-}$ on the left, and $k_{1}^{+}$ on the right.  Since these wavevectors have opposite signs, they have opposite norms, and form the only Hawking radiation channel.
\label{fig:dispersion_in_stationary_frame}}
\end{figure}

As remarked in Part I, waves of equal conserved frequency $\omega$ can be scattered into each other by an inhomogeneous flow, where $\omega$ is related to the frequency $\Omega$ in the rest frame of the fluid by the Doppler formula
\begin{equation}
\Omega^{2} = \left( \omega - u k \right)^{2} = c^{2}(k) k^{2} \,.
\label{eq:Doppler}
\end{equation}
The curves for each asymptotic region are shown in Figure \ref{fig:dispersion_in_stationary_frame}.  Restricting our attention to the counter-propagating branch of the dispersion curve (those waves which are right-moving with respect to the medium), Hawking radiation occurs when wavevectors of opposite sign couple to each other.  But when the asymptotic velocities are different as here, we must also pay attention to the signs of the group velocities of the various wavevectors, {\it i.e.} to their ingoing or outgoing character.  In Part I, there were two possible radiation channels; in the case considered here, one of the waves which could be emitted before can no longer be emitted because it only exists as an incoming wave, and hence we are left with only one radiation channel.  As we are considering a black hole horizon, this radiation is the analogue of the standard Hawking radiation: each member of the pair is long-wavelength, and they are emitted in opposite directions.

%%%%%%%%%%%%%%%%%%%%%%%

\subsection{Results}

\begin{figure}
\subfloat{\includegraphics[width=0.465\columnwidth]{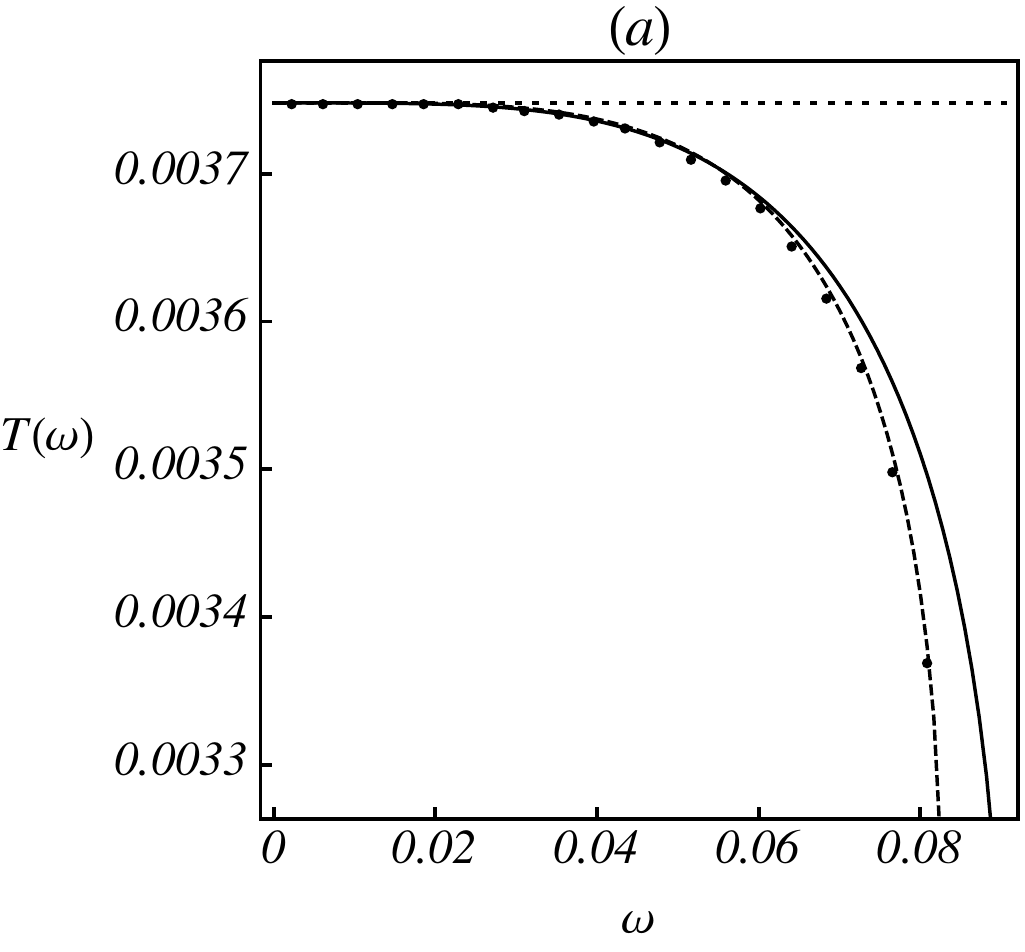} \, \includegraphics[width=0.45\columnwidth]{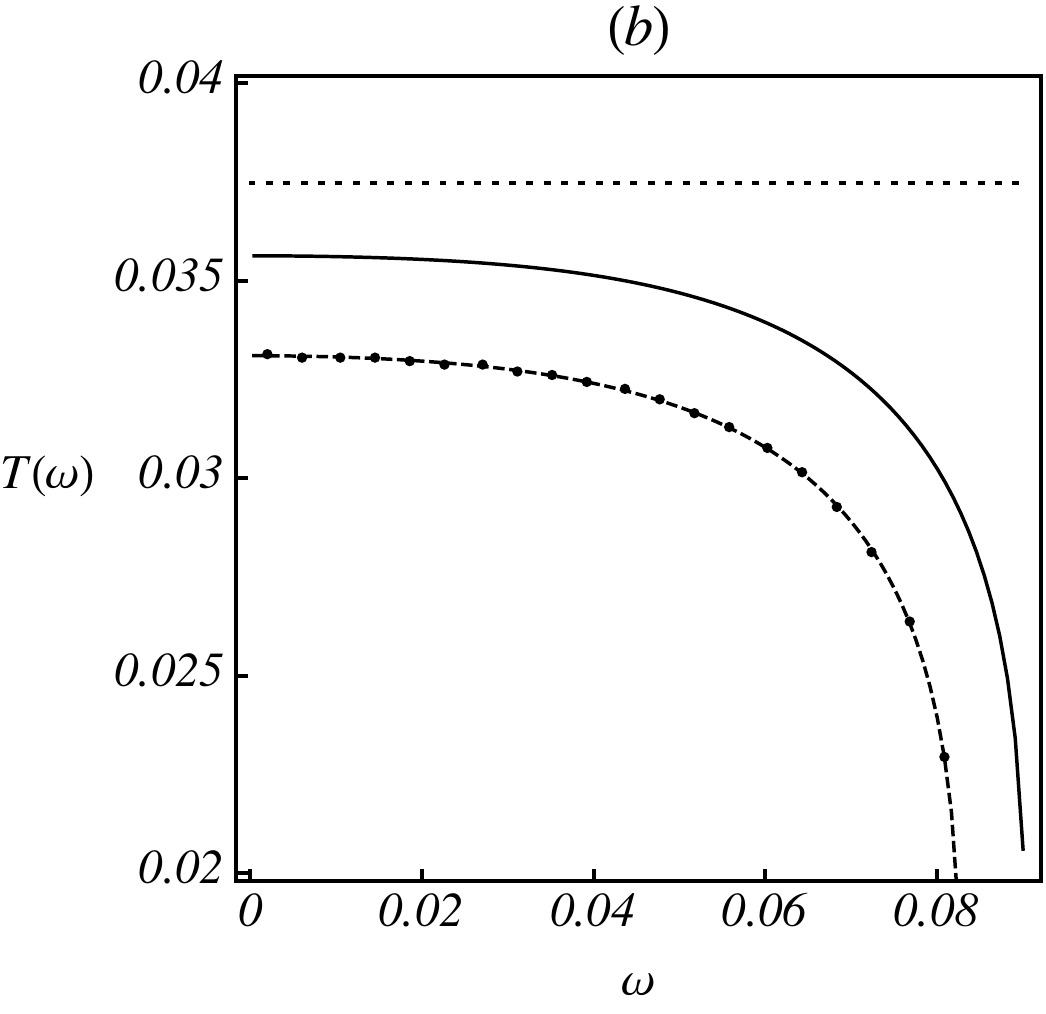}}
\caption{\textsc{Hawking spectra}: \, In $(a)$ are plotted the temperature (\ref{eq:temperature_spectrum}) of the Hawking spectra in the slowly-varying case, where $a = 0.118$; and in $(b)$ are shown the spectra in the rapidly-varying case, where $a=1.18$.  The solid curves plot the spectra for the high-order polynomial approximation to the first of dispersion relations (\ref{eq:model_dispersion_relations}), while the dashed curves show the spectra for the second of these dispersion relations.  Note that the spectra fall off to zero as $\omega$ approaches the maximum of the dispersion curve in Fig. \ref{fig:dispersion_in_stationary_frame}$(b)$, because above this frequency the member $k_{1}^{+}$ of the Hawking pair no longer exists.  The dotted lines shows the temperatures given by Hawking's orginal prediction (\ref{eq:Hawking_prediction}).  This is seen to be valid at low frequencies in the slowly-varying case, but overestimates the temperature in the rapidly-varying case, where the spectrum is seen to saturate in a manner that depends on the details of the dispersion profile.  The discrete points in these plots show the spectra for the low-degree polynomial dispersion as calculated by standard ODE methods, and are seen to agree very well with the results of the integral methods presented in this paper, especially in the rapidly-varying regime.
\label{fig:Hawking_spectra}}
\end{figure}

As in Part I, we plot the Hawking spectra using the frequency-dependent effective temperature
\begin{equation}
T(\omega) = \frac{\omega}{\mathrm{ln} \left( 1 + 1/\left| \beta_{\omega} \right|^{2} \right)} \,.
\label{eq:temperature_spectrum}
\end{equation}
The spectra for both the slowly-varying ($a=0.118$) and rapidly-varying ($a=1.18$) cases are shown in Figure \ref{fig:Hawking_spectra}.  Spectra for both of dispersion relations (\ref{eq:model_dispersion_relations}) are given, the first having been approximated by a high-degree polynomial as described in \S\ref{sub:Dispersion_relation}.  As we might have expected, the spectra are seen to agree with Hawking's original prediction in the slowly-varying case, but not in the rapidly-varying case; in the latter, the spectra are still well-described by a constant temperature at low frequencies, but this temperature is less than Hawking's prediction and depends on the details of the dispersion.  Also shown are the results of solving the position-space ODE for the low-degree polynomial dispersion; these are seen to agree very well with the results of the integral method presented here, demonstrating its validity.

\begin{figure}
\subfloat{\includegraphics[width=0.465\columnwidth]{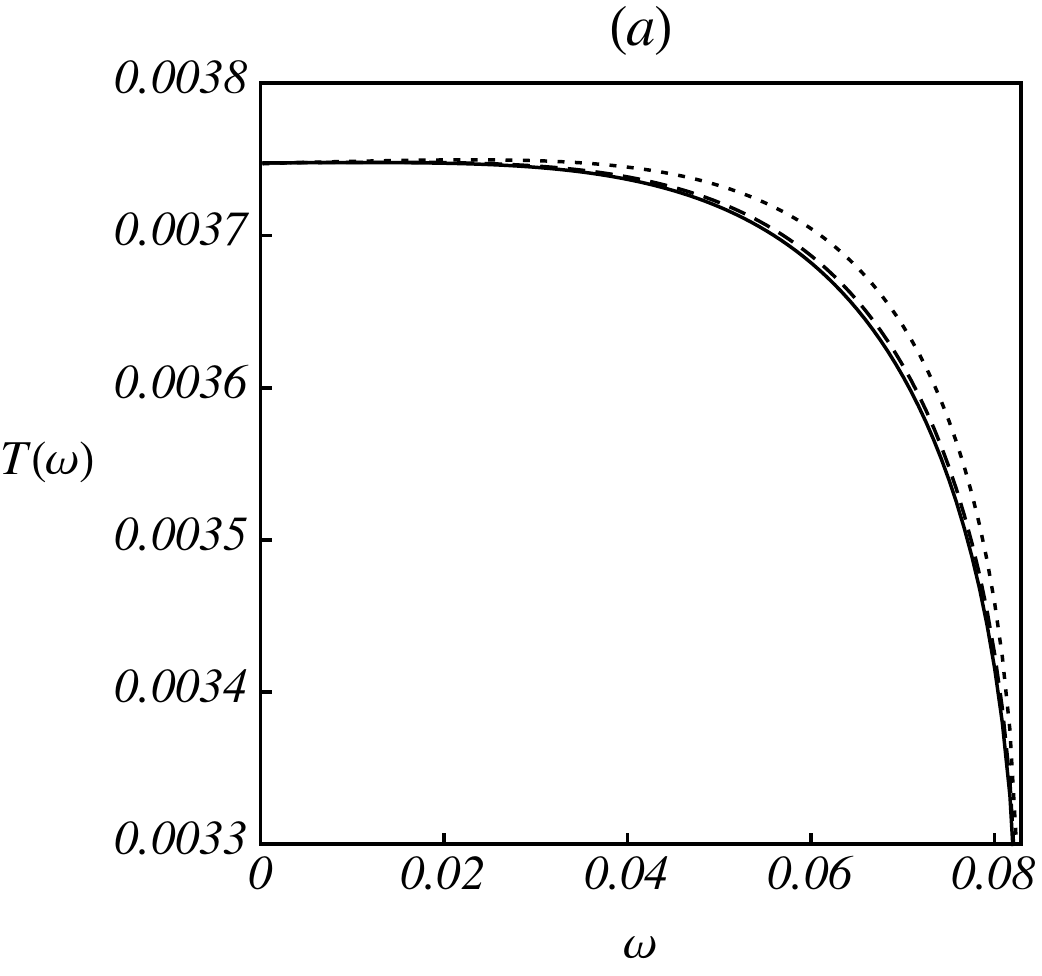} \, \includegraphics[width=0.45\columnwidth]{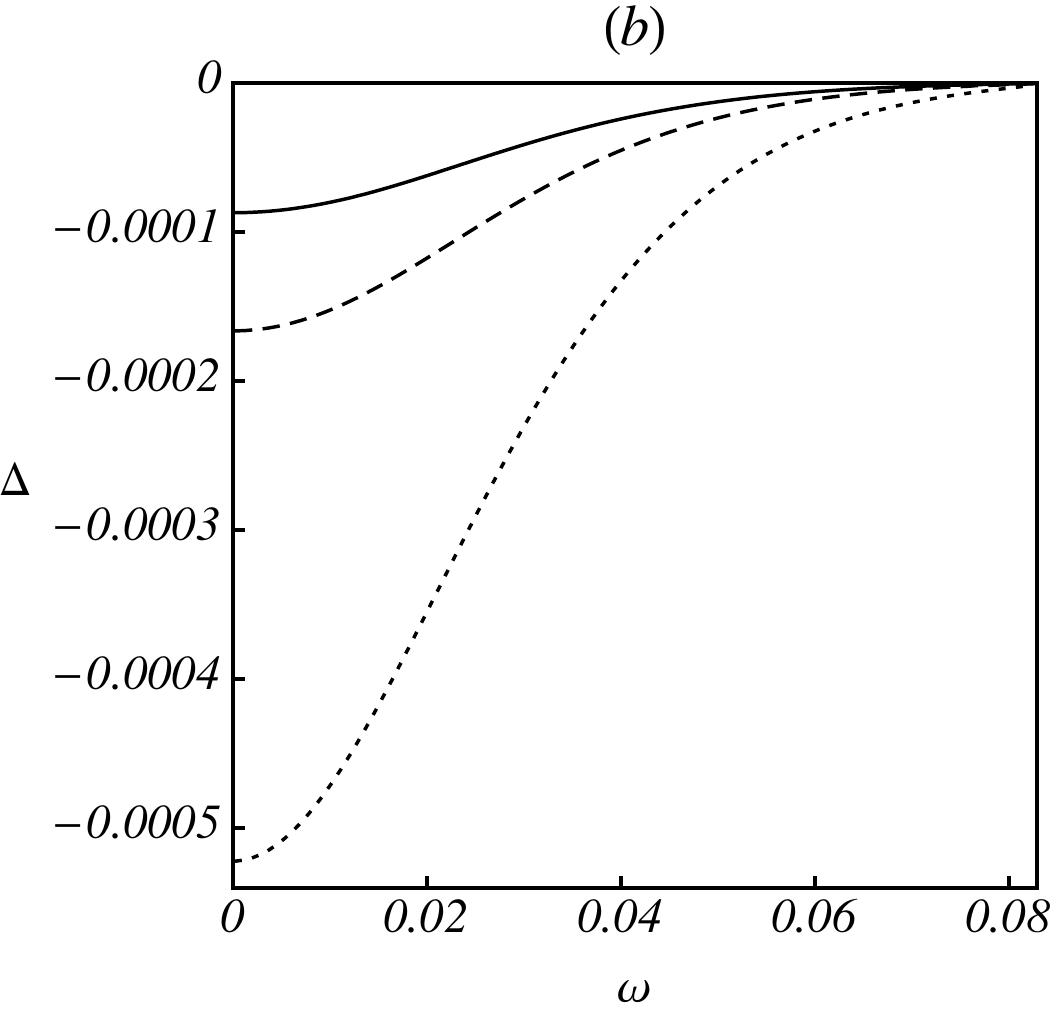}}
\caption{\textsc{Convergence for low-degree polynomial}: \, In $(a)$ are plotted, for a slowly-varying velocity profile ($a=0.118$) and with the dispersion relation given by the second of Eqs. (\ref{eq:model_dispersion_relations}), the spectra calculated with $M=100$ (dotted curve), $M=200$ (dashed curve) and $M=300$ (solid curve), where $M$ is the number of points in the discretised integration grid (see Appendix A).  In $(b)$, with the same correspondence between $M$ and the curve styles, is plotted the discrepancy in norm between the ingoing and outgoing waves.  This discrepancy, though it decreases in magnitude as $M$ increases, is large enough for the difference to be visible in the spectral temperature.
\label{fig:convergence_LP}}
\end{figure}

\begin{figure}
\subfloat{\includegraphics[width=0.5\columnwidth]{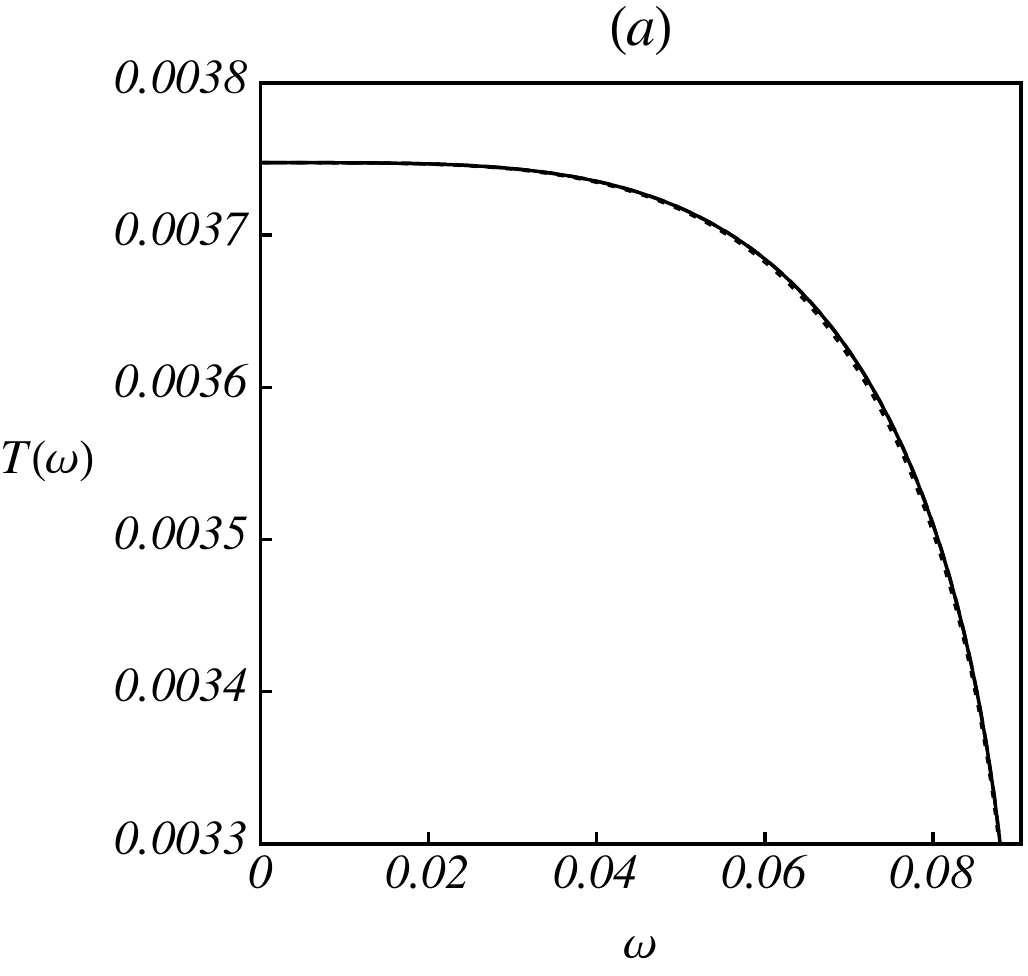} \, \includegraphics[width=0.45\columnwidth]{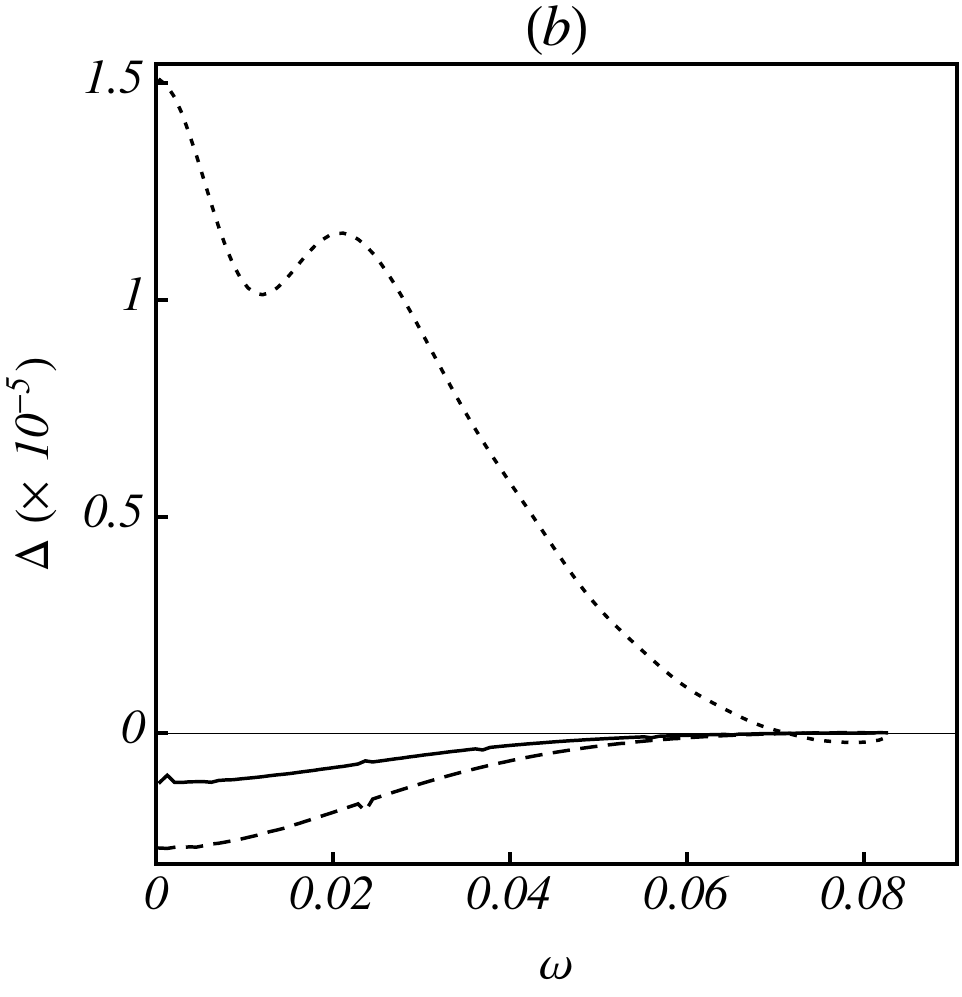}}
\caption{\textsc{Convergence for high-degree polynomial}: \, In $(a)$ are plotted, for a slowly-varying velocity profile ($a=0.118$) and with the dispersion relation given by a high-degree polynomial approximation of the first of Eqs. (\ref{eq:model_dispersion_relations}), the spectra calculated with $M=100$ (dotted curve), $M=200$ (dashed curve) and $M=300$ (solid curve), where $M$ is the number of points in the discretised integration grid.  In $(b)$ is shown the discrepancy in norm between the ingoing and outgoing waves.  This discrepancy is more than an order of magnitude smaller than the corresponding discrepancy for the low-degree polynomial dispersion (see Fig. \ref{fig:convergence_LP}$(b)$), and its effect on the spectral temperature is small.
\label{fig:convergence_HP}}
\end{figure}

In Figures \ref{fig:convergence_LP} and \ref{fig:convergence_HP} is shown how the calculated spectra converge for the slowly-varying velocity profile as the number of points $M$ in the discretised integration grid increases, with $M$ taking the values $100$, $200$ and $300$.  (The scaling parameter $k_{0}$ is fixed at $2$; see Appendix A for its definition.)  Figure \ref{fig:convergence_LP}$(a)$ shows the spectra for the low-degree polynomial dispersion (the second of dispersion relations (\ref{eq:model_dispersion_relations})), while in Figure \ref{fig:convergence_LP}$(b)$ is shown the overall discrepancy $\Delta$ in the norm between ingoing and outgoing solutions:
\begin{equation}
\Delta_{\omega} = \sum_{j} \left| \alpha_{\omega,j} \right|^{2} - \sum_{j} \left| \beta_{\omega,j} \right|^{2} - 1 \,,
\label{eq:Delta}
\end{equation}
where $\alpha_{\omega,j}$ and $\beta_{\omega,j}$ are the scattering amplitudes into modes of the same and opposite norm, respectively.  (See \S II of Part I for a description of norm and its conservation.)  For an exact solution, this must be zero, and we see that it decreases in magnitude as $M$ is increased.  Figure \ref{fig:convergence_HP} shows the same information for the high-degree polynomial approximation of the first of dispersion relations (\ref{eq:model_dispersion_relations}).  Interestingly, the solution for the high-degree polynomial converges faster than that for the low-degree polynomial, whose discrepancy is visible in the spectrum and explains the slight mismatch seen with the results of the ODE solution in Fig. \ref{fig:Hawking_spectra}$(a)$.  This trend occurs also for the rapidly-varying velocity profile, though both values of $\Delta$ are an order of magnitude smaller than their counterparts in Figs. \ref{fig:convergence_LP}$(b)$ and \ref{fig:convergence_HP}$(b)$, while the differences in the spectra for different $M$ are too small to discern, much like in Fig. \ref{fig:convergence_HP}$(a)$.

%%%%%%%%%%%%%%%%%%%%%%%%%%%%%%%%%%%%%%%%%%%%%%%%%%%

\section{Conclusion
\label{sec:Conclusion}}

In a previous paper (Part I), it was shown that the scattering of dispersive waves in one spatial dimension can be solved using an integral method in Fourier space, which is more generally applicable than the standard methods of directly solving an ODE in position space.  There, however, the method was restricted to situations in which the background was the same in both the left- and right-hand asymptotic regions.  The current paper generalises this integral method to cases in which the asymptotic backgrounds are not symmetric.  As before, the method utilises the analyticity properties of the half-Fourier transforms of the solution.  In order to reduce the calculational complexity of the method, the dispersion relation has been restricted to a polynomial, but no restriction has been placed on the degree of this polynomial, allowing an arbitrary dispersion relation to be approximated by a polynomial within a given region of interest.  Increasing the degree of this polynomial will slow the calculation as it will increase the number of complex wavevector solutions, but this is typically much less than the number of points on the discretised integration grid, and it is the latter that is dominant in determining the required amount of computation.  When the degree of the polynomial dispersion relation is small, solution of the position-space ODE is possible, and the integral method has been shown to agree with the ODE solution in such a case.

It is also to be noted that, taking the difference in norm between incoming and outgoing waves as a measure of the accuracy of the numerical method, the results for the asymmetric velocity profile considered in Section \ref{sec:Application} are much less accurate than the results for the symmetric velocity profile of Part I, with comparable model and discretisation parameters.  As we have seen, asymmetry of the asymptotic regions substantially complicates the details of the integral method, and it is quite likely that the numerical scheme adopted here (as described in the Appendices) is not the optimal one.  How numerical optimisation might proceed, however, is unclear at the present time.

%%%%%%%%%%%%%%%%%%%%%%%%%%%%%%%%%%%%%%%%%%%%%%%%%%%

\begin{acknowledgments}
I wish to thank Ulf Leonhardt for helpful discussions and guidance on the work whose results are presented here.  I also thank South China Normal University and the Weizmann Institute of Science for their hospitality and financial support during the period in which this work was done.
\end{acknowledgments}

%%%%%%%%%%%%%%%%%%%%%%%%%%%%%%%%%%%%%%%%%%%%%%%%%%%

\newpage
\appendix
\appendixpage
\numberwithin{equation}{section}

\section{Change of variables for numerical integration
\label{app:Change_of_variables_for_numerical_integration}}

The change of variables implemented for the purpose of numerical integration is given by the equivalent formulae (for $\zeta \in (-1,1)$)
\begin{alignat}{3}
\frac{k}{k_{0}} = \frac{\zeta}{1-\zeta^{2}} & \qquad \Longleftrightarrow & \qquad \zeta = \frac{-1+\sqrt{1+4\left(k/k_{0}\right)^{2}}}{2\left(k/k_{0}\right)} \,.
\label{eq:change_of_variables}
\end{alignat}
Around $k=0$ or $\zeta=0$, the relation is linear, so that the change of variables corresponds simply to a rescaling of $k$ with scaling parameter $k_{0}$.  As $\zeta \rightarrow \pm 1$, the corresponding value of $k$ diverges, and the infinite integration range $(-\infty,+\infty)$ is thus mapped onto the finite interval $(-1,+1)$.  In the transformation of integrals, we must also multiply by the Jacobian $\mathrm{d}k/\mathrm{d}\zeta$, where
\begin{equation}
\frac{\mathrm{d}k}{\mathrm{d}\zeta} = \frac{k_{0}}{2} \left( \frac{1}{\left(1-\zeta\right)^{2}} + \frac{1}{\left(1+\zeta\right)^{2}} \right) \,.
\label{eq:dkdzeta}
\end{equation}
As $k \rightarrow \pm \infty$ or $\zeta \rightarrow \pm 1$, we have $\mathrm{d}k/\mathrm{d}\zeta \rightarrow \left(2/k_{0}\right) \times k^{2}$, so multiplication of this factor by a product of two function each of which behaves asymptotically like $1/k$ results in a finite limiting value as $\zeta \rightarrow \pm 1$.  It is convenient to distribute the factor $\mathrm{d}k/\mathrm{d}\zeta$ equally between the two factors; that is, each function appearing in the integrand, when the change of variable from $k$ to $\zeta$ is effected, is multiplied by $\sqrt{\mathrm{d}k/\mathrm{d}\zeta}$.  Those functions with a $1/k$ asymptotic behaviour thus approach a finite limiting value as $\zeta \rightarrow \pm 1$, so that each term of the integrand is well-behaved at the boundaries of the integral over $\zeta$.

It is important to check that the inverse kernel $V(k,k^{\prime})$ is also well-behaved after the change of variables (\ref{eq:change_of_variables}), and that it still corresponds to a straightforward matrix inverse in its discretised form.  Indeed, using the following equation for the change of variables of the $\delta$ function:
\begin{equation}
\delta\left(k(\zeta)-k(\zeta^{\prime})\right) = \left[ \frac{\mathrm{d}k}{\mathrm{d}\zeta}\left(\zeta\right) \, \frac{\mathrm{d}k}{\mathrm{d}\zeta}\left(\zeta^{\prime}\right) \right]^{-1/2} \, \delta(\zeta-\zeta^{\prime}) \,,
\end{equation}
it is straightforward to show that the defining equation for the transformed inverse kernel becomes
\begin{equation}
\int_{-1}^{1} \mathrm{d}\zeta^{\prime\prime} \, V(\zeta,\zeta^{\prime\prime}) \, \left[ \delta(\zeta^{\prime\prime} - \zeta^{\prime}) + \bar{\kappa}(\zeta^{\prime\prime},\zeta^{\prime}) \right] = \delta(\zeta-\zeta^{\prime})
\label{eq:inverse_kernel_zeta}
\end{equation}
where
\begin{subequations}\begin{eqnarray}
V(\zeta,\zeta^{\prime\prime}) & = & \sqrt{\frac{\mathrm{d}k}{\mathrm{d}\zeta}(\zeta)} \,\, V\left( k(\zeta), k(\zeta^{\prime\prime}) \right) \, \sqrt{\frac{\mathrm{d}k}{\mathrm{d}\zeta}(\zeta^{\prime\prime})} \,, \\
\bar{\kappa}(\zeta^{\prime\prime},\zeta^{\prime}) & = & \sqrt{\frac{\mathrm{d}k}{\mathrm{d}\zeta}(\zeta^{\prime\prime})} \,\, \bar{\kappa}\left( k(\zeta^{\prime\prime}), k(\zeta^{\prime}) \right) \, \sqrt{\frac{\mathrm{d}k}{\mathrm{d}\zeta}(\zeta^{\prime})} \,.
\end{eqnarray}\end{subequations}
So the transformed $V$, including left- and right-multiplication by $\sqrt{\mathrm{d}k/\mathrm{d}\zeta}$, is itself an inverse kernel and can be calculated by discretising the interval $(-1,+1)$ of $\zeta$ with uniform spacing $\Delta\zeta$.

The change of variables described in this appendix copes well with the $1/k$ behaviour of the functions to be integrated, so long as we are careful to choose $k_{0}$ such that no non-trivial structures in the integrand are pushed towards the boundaries at $\zeta = \pm 1$, since this would squeeze them into a narrow region of $\zeta$-space and require a very small spacing $\Delta \zeta$ to be integrated accurately.

%%%%%%%%%%%%%%%%%%%%%%%%%%%%%%%%%%%%%%%%%%%%%%%%%%%

\section{Matrix multiplication suitable for numerics
\label{app:Matrix_multiplication_suitable_for_numerics}}

Here we shall say a few words regarding the numerical implementation of Eqs. (\ref{eq:matrix_elements_exact}), or more precisely of their discretised versions (\ref{eq:matrix_elements_discretised}).  Although these equations are substantially more complicated than their counterparts in Part I, the considerations of Appendix B of Part I still apply.  In particular, when calculating a spectrum over which $\omega$ varies, we calculate the large $\omega$-independent matrices of order $M^{2}$ (where $M$ is the number of points in the discretised integration grid) at the outset, before calculation of the spectrum itself.  These $M^{2}$-sized matrices are the left and right components of $K(k,k^{\prime})$, decomposed into coefficients of powers of $\omega$.  In the Unruh model with the kernel of Eq. (\ref{eq:unruh_kernel}), $K(k,k^{\prime})$ can be considered as a first-degree polynomial in $\omega$, so we can write $K(k,k^{\prime}) = K^{(0)}(k,k^{\prime}) + \omega \, K^{(1)}(k,k^{\prime})$, and upon splitting into left and right components (see Appendix A of Part I) we have
\begin{equation}
K_{\sigma}(k,k^{\prime}) = K^{(0)}_{\sigma}(k,k^{\prime}) + \omega \, K^{(1)}_{\sigma}(k,k^{\prime}) \,.
\end{equation}
There are thus four $\omega$-independent $M \times M$ matrices to be stored before calculation of the $\omega$-dependent $\mathcal{S}$-matrix.  The $\omega$-dependent matrices of order $M \times M$ are constructed from these using standard matrix manipulations, while those of order $M$ or less are small enough to be calculated directly for each value of $\omega$.

\subsection{Number of matrices and notation}

The complications that arise due to asymmetric asymptotics are not in the required manipulations themselves, but simply in the number of matrices that we require to define.  This is mainly due to there being two different sets of solutions of the dispersion relation, but it leads to more than a straightforward doubling of the number of matrices.  As we have seen in Eqs. (\ref{eq:general_split_int_eqns}), it necessitates the introduction of a complete set of polynomials, and hence introduces the $\mathcal{P}^{\rho}$ matrices into the final expression (\ref{eq:step_scattering_matrix}) for the scattering matrix.  It also induces two types of ``overbar'' transformation (described in Eq. (\ref{eq:overbar_defn})) corresponding to the two different $\tilde{g}_{\sigma}(k)$ functions, with the functions $\bar{\kappa}_{\rho,j}(k^{\prime})$ appearing in the integrals of Eqs. (\ref{eq:matrix_elements_exact}) being sums of different types.  A final cause of the increase in the number of matrices is the appearance of the complex roots of the dispersion relations in Eqs. (\ref{eq:kappaLR_defn})-(\ref{eq:general_kappa_defns}), requiring a further set of matrices in which these complex roots appear as arguments.

This large number of matrices can be handled on adopting an economical system of notation.  As in Part I, we use a hat to denote a matrix to be stored or calculated numerically, and we use labels to indicate which wavevectors appear as arguments in a given function.  There are three distinct groups of such wavevectors:
\begin{itemize}
\item the elements of the discretised integration grid, indicated by the label $\delta$;
\item the real roots of the dispersion relations, indicated by the label $r$; and
\item the complex roots of the dispersion relations, indicated by the label $c$.
\end{itemize}
The latter two groups also require labelling by $\sigma$, to distinguish the left and right solution sets.  However, $\sigma$ labels also appear on the functions and matrices themselves, and in the majority of cases these correspond exactly to the $\sigma$ labels on the arguments.  In those cases where the labels on the functions and arguments do not coincide, we indicate the difference with a minus sign in front of the label $r$ or $c$.  This notation allows us to represent most of the matrices required in the numerical algorithm.

\subsection{Numerical formation of matrices}

We need to form matrices of the $\kappa(k,k^{\prime})$ functions from those of the original kernel $K(k,k^{\prime})$.  From Eq. (\ref{eq:general_kappa_different_sigma}) we see that, for different $\sigma$ labels, $\kappa^{-\sigma}_{\quad\sigma}(k,k^{\prime})$ can be formed from matrix multiplication which sums over the complex roots; and, from Eqs. (\ref{eq:kappaLR_defn}) and (\ref{eq:general_kappa_split}), the corresponding function with equal $\sigma$ labels, $\kappa^{\sigma}_{\,\,\sigma}(k,k^{\prime})$, is most easily formed by sutracting $\kappa^{-\sigma}_{\quad\sigma}(k,k^{\prime})$ from $F^{-1}(k) \cdot K_{\sigma}(k,k^{\prime})$, where $F(k)$ is the polynomial whose roots are the forbidden wavevectors.  In matrix form, we define
\begin{subequations}\begin{eqnarray}
\left[ \hat{K}^{\sigma c}_{\,\,\,\,\sigma \delta} \right]_{qm} & = & K_{\sigma}\left(k_{\sigma,q}^{c\star},k_{m}\right) \,, \\
\left[ \hat{\Delta}^{\delta}_{\,\,\sigma c} \right]_{nq} & = & \frac{1}{\left(k_{n}-k_{\sigma,q}^{c\star}\right) \, C_{\sigma,q}} \,, \label{eq:Delta_c}\\
\left[ \hat{F} \right]_{nm} & = & \prod_{q=1}^{N_{L}^{c}} \left( k_{n}-k_{L,q}^{c\star} \right) \, \prod_{q=1}^{N_{R}^{c}} \left(k_{n}-k_{R,q}^{c\star} \right) \, \delta_{nm} \,,
\end{eqnarray}\end{subequations}
from which we form the $\hat{\kappa}$ matrices via the equations
\begin{subequations}\begin{eqnarray}
\hat{\kappa}^{-\sigma \delta}_{\quad\sigma \delta} & = & \hat{\Delta}^{\delta}_{\,\,\sigma c} \cdot \hat{K}^{\sigma c}_{\,\,\,\,\sigma \delta} \,, \\
\hat{\kappa}^{\sigma \delta}_{\,\,\,\,\sigma \delta} & = & \hat{F}^{-1} \cdot \hat{K}^{\delta}_{\,\,\sigma \delta} - \hat{\kappa}^{-\sigma \delta}_{\quad\sigma \delta} \,. \label{eq:kappa_matrix}
\end{eqnarray}\end{subequations}
Matrix equations entirely analogous to these allow us to calculate all the $\hat{\kappa}$ matrices, with the labels $\delta$ exchanged for $r$ and $c$.  (Note that, when the upper label is of $r$-type rather than $\delta$-type, the subtracted matrix in Eq. (\ref{eq:kappa_matrix}) will have an upper label of $-r$-type because the first arguments are $k_{\sigma,j}^{r}$ rather than $k_{-\sigma,j}^{r}$.)  One of these is exactly the first of the matrices appearing in the sums of Eqs. (\ref{eq:matrix_elements_discretised}): this is the discretised version of the function $\kappa_{i}^{\sigma}(k) = \kappa^{\sigma}\left(k_{\sigma,i}^{r},k\right) = \kappa^{\sigma}_{\,\,L}\left(k_{\sigma,i}^{r},k\right)+\kappa^{\sigma}_{\,\,R}\left(k_{\sigma,i}^{r},k\right)$, and in the notation adopted here is the $N_{\sigma} \times M$ matrix sum $\hat{\kappa}^{\sigma r}_{\,\,L \delta} + \hat{\kappa}^{\sigma r}_{\,\,R \delta}$.

We next need an efficient way of performing the transformation denoted by the overbar.  This has already been described in Appendix B of Part I: in the case of the matrices with only $\delta$ labels, we need
\begin{subequations}\begin{eqnarray}
\left[ \hat{\Delta}^{\delta}_{\,\,\sigma r} \right]_{nj} & = & \frac{1}{\left( k_{n} - k_{\sigma,j}^{r} \right) \, \tilde{g}_{\sigma}^{\prime}\left( k_{\sigma,j}^{r} \right)} \,, \\
\left[ \hat{\tilde{g}}_{\sigma \delta} \right]_{nm} & = & \tilde{g}_{\sigma}\left(k_{n}\right) \, \delta_{nm} \,.
\end{eqnarray}\end{subequations}
Then, having already defined the $\kappa$ matrices, we have
\begin{equation}
\hat{\bar{\kappa}}^{\sigma \delta}_{\,\,\,\,\sigma^{\prime} \delta} = \hat{\tilde{g}}_{\sigma \delta}^{-1} \cdot \hat{\kappa}^{\sigma \delta}_{\,\,\,\,\sigma^{\prime} \delta} - \hat{\Delta}^{\delta}_{\,\,\sigma r} \cdot \hat{\kappa}^{\sigma r}_{\,\,\,\,\sigma^{\prime} \delta} \,.
\end{equation}
Again, the various $\hat{\bar{\kappa}}$ matrices are defined analogously, simply by changing the $\delta$ labels for $r$ and $c$ labels.  The final factor appearing in the sums of the first two of Eqs. (\ref{eq:matrix_elements_discretised}) are within this group: they are the discretised versions of $\bar{\kappa}_{\sigma,j}(k) = \bar{\kappa}_{\sigma}\left(k,k_{\sigma,j}^{r}\right) = \bar{\kappa}^{L}_{\,\,\sigma}\left(k,k_{\sigma,j}^{r}\right) + \bar{\kappa}^{R}_{\,\,\sigma}\left(k,k_{\sigma,j}^{r}\right)$, denoted here by the $M \times N_{\sigma}$ matrix sum $\hat{\bar{\kappa}}^{L \delta}_{\,\,\,\,\sigma r} + \hat{\bar{\kappa}}^{R \delta}_{\,\,\,\,\sigma r}$.

To perform integrals over $\zeta$, we require a diagonal matrix whose elements are the values of $\sqrt{\mathrm{d}k/\mathrm{d}\zeta}$:
\begin{equation}
\left[ \hat{j} \right]_{nm} = \sqrt{\frac{\mathrm{d}k}{\mathrm{d}\zeta}\left(\zeta_{n}\right)} \,\, \delta_{nm} \,.
\end{equation}
This allows us to find $\hat{V}$, the discretised form of the transformed inverse kernel defined by Eq. (\ref{eq:inverse_kernel_zeta}): it is given by
\begin{equation}
\hat{V} = \left[ \mathbb{1}_{M} + \Delta \zeta \sum_{\sigma} \sum_{\sigma^{\prime}} \hat{j} \cdot \hat{\bar{\kappa}}^{\sigma \delta}_{\,\,\,\,\sigma^{\prime} \delta} \cdot \hat{j} \right]^{-1} \,.
\end{equation}
The matrices $\mathcal{M}^{\sigma}_{\,\,\sigma^{\prime}}$ of Eq. (\ref{eq:LR_matrix_equation}) may now be written in terms of matrices defined above:
\begin{subequations}\begin{eqnarray}
\mathcal{M}^{\sigma}_{\,\,\sigma} & = & \frac{s_{\sigma}}{2\pi i} \hat{\tilde{g}}^{\prime}_{\sigma r} + \hat{\kappa}^{\sigma r}_{\,\,\,\,\sigma r} - \Delta \zeta \cdot \left( \hat{\kappa}^{\sigma r}_{\,\,L \delta} + \hat{\kappa}^{\sigma r}_{\,\,R \delta} \right) \cdot \hat{j} \cdot \hat{V} \cdot \hat{j} \cdot \left( \hat{\bar{\kappa}}^{L \delta}_{\,\,\,\,\sigma r} + \hat{\bar{\kappa}}^{R \delta}_{\,\,\,\,\sigma r} \right) \,, \\
\mathcal{M}^{\sigma}_{\,\,-\sigma} & = & \hat{\kappa}^{\sigma r}_{\,\,-\sigma r} - \Delta \zeta \cdot \left( \hat{\kappa}^{\sigma r}_{\,\,L \delta} + \hat{\kappa}^{\sigma r}_{\,\,R \delta} \right) \cdot \hat{j} \cdot \hat{V} \cdot \hat{j} \cdot \left( \hat{\bar{\kappa}}^{L \delta}_{\,\,-\sigma r} + \hat{\bar{\kappa}}^{R \delta}_{\,\,-\sigma r} \right) \,.
\end{eqnarray}\end{subequations}

The $\mathcal{P}^{\sigma}$ matrices are somewhat simpler.  The elements of $\hat{p}^{\sigma r}$ are straightforward evaluations of the $p_{j}(k)$ at the roots of the dispersion relation in asymptotic region $\sigma$:
\begin{equation}
\left[ \hat{p}^{\sigma r} \right]_{ij} = p_{j}\left( k_{\sigma,i}^{r} \right) \,.
\end{equation}
Also, since $p_{j}(k)$ is a polynomial of degree $N-1$ and $\tilde{g}_{\sigma}(k)$ is a rational function that behaves asymptotically like $k^{N}$, $p_{j}(k)/\tilde{g}_{\sigma}(k)$ behaves asymptotically like $1/k$ and is exactly a sum over the poles at the zeros of $\tilde{g}_{\sigma}(k)$.  When constructing $\bar{p}^{\sigma}_{j}(k)$, the poles on the real axis are subtracted, so $\bar{p}^{\sigma}_{j}(k)$ is exactly the sum over the poles at the {\it complex} roots of $\tilde{g}_{\sigma}(k)$, and $\bar{p}_{j}(k)$ is the difference between these sums for $\sigma=R$ and $\sigma=L$:
\begin{equation}
\bar{p}_{j}(k) = \sum_{q=1}^{N_{R}^{c}} \frac{p_{j}\left(k_{R,q}^{c}\right)}{\left(k-k_{R,q}^{c}\right) \tilde{g}^{\prime}_{R}\left(k_{R,q}^{c}\right)} - \sum_{q=1}^{N_{L}^{c}} \frac{p_{j}\left(k_{L,q}^{c}\right)}{\left(k-k_{L,q}^{c}\right) \tilde{g}^{\prime}_{L}\left(k_{L,q}^{c}\right)} \,.
\end{equation}
Discretisation is achieved by defining the matrices
\begin{subequations}\begin{eqnarray}
\left[ \hat{p}^{\sigma c} \right]_{qj} & = & p_{j}\left(k_{\sigma,q}^{c}\right) \,,\\
\left[ \hat{\Delta}^{\delta}_{\,\,\sigma cg} \right]_{nq} & = & \frac{1}{\left(k_{n} - k_{\sigma,q}^{c}\right) \tilde{g}^{\prime}_{\sigma}\left(k_{\sigma,q}^{c}\right)} \,, \label{eq:Delta_cg}
\end{eqnarray}\end{subequations}
where we have added a $g$ subscript in Eq. (\ref{eq:Delta_cg}) to distinguish it from Eq. (\ref{eq:Delta_c}).  From these we can form
\begin{equation}
\hat{\bar{p}} = \hat{\Delta}^{\delta}_{\,\,L cg} \cdot \hat{p}^{L c} - \hat{\Delta}^{\delta}_{\,\,R cg} \cdot \hat{p}^{R c} \,.
\end{equation}
The $\mathcal{P}^{\sigma}$ matrices are then given by
\begin{equation}
\hat{\mathcal{P}}^{\sigma} = \hat{p}^{\sigma r} - \Delta\zeta \cdot \left( \hat{\kappa}^{\sigma r}_{\,\,L \delta} + \hat{\kappa}^{\sigma r}_{\,\,R \delta} \right) \cdot \hat{j} \cdot \hat{V} \cdot \hat{j} \cdot \hat{\bar{p}} \,.
\end{equation}

%%%%%%%%%%%%%%%%%%%%%%%%%%%%%%%%%%%%%%%%%%%%%%%%%%%

%%%%%%%%%%%%%%%%%%%%%%%%%%%%%%%%%%%%%%%%%%%%%%%%%

\newpage

%%%%%%%%%%%%%%%

%%%%%%%%%%%%%%%

%%%%%%%%%%%%%%%

%%%%%%%%%%%%%%%

%%%%%%%%%%%%%%%

%%%%%%%%%%%%%%%

\end{document}